\documentclass[11pt,a4paper]{article}
\usepackage[utf8]{inputenc}
\usepackage[T1]{fontenc}
\usepackage{amsmath}
\usepackage{amsfonts}
\usepackage{graphicx}
\usepackage{caption}
\usepackage{booktabs}
\usepackage{multirow}
\usepackage{siunitx}
\usepackage{amssymb}
\usepackage{color}
\usepackage{colortbl}
\usepackage{siunitx}
\usepackage{booktabs}
\usepackage{longtable}
\usepackage[left=2.50cm, right=2.50cm, top=2.50cm, bottom=2.50cm]{geometry}
\usepackage[onehalfspacing]{setspace}
\usepackage{placeins}
\usepackage{authblk}
\usepackage{lipsum}

\usepackage{apacite} 				
\usepackage{natbib}

\title{Flexible domain prediction using mixed effects random forests}
\author[*]{Patrick Krennmair}
\author[**]{Timo Schmid}
\affil[*]{\small{Institute of Statistics and Econometrics, Freie Universit\"{a}t Berlin, Berlin, Germany}}
\affil[**]{\small{Institute of Statistics, Otto-Friedrich-Universit\"{a}t Bamberg, Bamberg, Germany}}
\date{}

\begin{document}

\maketitle
\onehalfspacing
\normalsize
\setlength{\parindent}{0em}

\begin{abstract}
This paper promotes the use of random forests as versatile tools for estimating spatially disaggregated indicators in the presence of small area-specific sample sizes. Small area estimators are predominantly conceptualized within the regression-setting and rely on linear mixed models to account for the hierarchical structure of the survey data. In contrast, machine learning methods offer non-linear and non-parametric alternatives, combining excellent predictive performance and a reduced risk of model-misspecification. Mixed effects random forests combine advantages of regression forests with the ability to model hierarchical dependencies. This paper provides a coherent framework based on mixed effects random forests for estimating small area averages and proposes a non-parametric bootstrap estimator for assessing the uncertainty of the estimates. We illustrate advantages of our proposed methodology using Mexican income-data from the state Nuevo León. Finally, the methodology is evaluated in model-based and design-based simulations comparing the proposed methodology to traditional regression-based approaches for estimating small area averages.
\end{abstract}


\section{Introduction}\label{sec:1}
Having accurate and detailed information on social and economic conditions, summarised by appropriate indicators, is imperative for the efficient implementation of policies. The term \textit{detailed} is used to signify information that extends beyond aggregate levels into highly disaggregated geographical and other domains (e.g. demographic groups). The term \textit{accurate} refers to information that is estimated with appropriate level of precision and is comparable over space and time. Simply analysing data from national sample surveys is not enough for achieving the dual target of accurate and detailed information. This is mainly due to the reduction of sample sizes as the level of required detail increases. The achievement of this dual target demands appropriate model-based methodology collectively referred to as Small Area Estimation (SAE).

SAE-methods can be broadly divided in two classes: first, area-level models \citep{Fay_Heriot1979} assume that only aggregated data for the survey and for the auxiliary information is available. Second, unit-level models \citep{Battese_etal1988} - further labelled as BHF - require access to the survey and to the auxiliary information on micro-level. A versatile extension of the BHF model is the EBP-approach by \citet{Molina_rao2010}. The EBP is capable to estimate area-level means as well as other linear and non-linear indicators. Both classes (area-level and unit-level models) are predominantly regression-based models, where the hierarchical structure of observations is modelled by random effects. These linear mixed models (LMM) assume normality of random effects and error terms. Focusing on social and economic inequality data, the required assumptions for LMMs hardly meet empirical evidence. \citet{JiangRao2020} remind, that optimality results and predictive performance in model-based SAE are inevitably connected to the validity of model assumptions. Without theoretical and practical considerations regarding improperly met assumptions, estimates are potentially biased and mean squared error (MSE) estimates unreliable.

One strategy to prevent model-failure, is the assurance of normality by transforming the dependent variable \citep{sugasawa2017transforming,Rojas_etal2019}. For instance, \cite{Rojas_etal2019} generalize the EBP with a data-driven transformation on the dependent variable, such that normality assumptions can be met in transformed settings. Further details on how to obtain the most-likely transformation parameter that improves the performance of unit-level models are available in \cite{Rojas_etal2019} and \cite{sugasawa2019adaptively} or from a more applied perspective in \cite{Tzavidis_etal2018}. Apart from transformation strategies, another alternative is the use of models with less restrictive (parametric) assumptions to avoid model-failure. For instance, \cite{DialloRao2018} and \cite{Graf_etal2019} formulate the EBP under more flexible distributional assumptions. \textcolor{black}{Alternatively, \citet{Cha06} propose an approach for estimating area-level means based on M-quantile models, which are a robust method avoiding distributional assumptions of LMMs including the formal specification of area-level random effects. \citet{Tzavidis_etal2010} and \citet{Marchetti_Tzavidis2021} extended this approach to allow for estimating more complex statistics, like quantiles of area-specific distribution functions and non-linear indicators.} Semi- or non-parametric approaches for the estimation of area-level means were investigated among others by \citet{Opsomeretal2008}. \citet{Opsomeretal2008} use penalized splines regression, treating the coefficients of spline components as additional random effects within the LMM setting.

A distinct methodological option to avoid the parametric assumptions of LMMs are the class of machine learning methods. These methods are not only limited to parametric models and `learn' predictive relations from data, including higher order interactions between covariates, without explicit model assumptions \citep{Hastie_etal2009, Varian2014}. Among the broad class of machine learning methods, we focus on tree-based models and particularly on random forests \citep{Breiman2001} because they exhibit excellent predictive performance in the presence of outliers and implicitly solve problems of model-selection \citep{Biau_Scornet2016}. In general, the predictive perspective of (tree-based) machine learning methods conceptually transfers straight forward to the methodology of unit-level SAE-models: survey data is used to construct a model with predictive covariates. Subsequently, auxiliary information from a supplementary data source (usually census, register or administrative data) is utilized to obtain predictions over sampled and non-sampled areas. From a machine learning perspective, the survey data serves as an implicit training set to construct a proper model, while supplementary data is used to predict final indicators. Nevertheless, \citet{JiangRao2020} claim, that results from machine learning methods in SAE are harder to be interpreted and justified by SAE-practitioners, compared to LMM-alternatives.

We aim to fill this gap by providing a consistent framework enabling a coherent use of tree-based machine learning methods in SAE. In particular, we \textcolor{black}{incorporate random forests within the methodological tradition of SAE by proposing a non-linear, data-driven, and semi-parametric alternative for the estimation of area-level means by using Mixed Effects Random Forests (MERF) \citep{Hajjem2014}. We focus on the construction of area-level mean-estimates using MERFs for sampled and out-of-sample domains. Our proposed model-based estimator consists of a composite model of a structural component, accounting for hierarchical dependencies of survey data with random effects and a non-parametric random forest, which models fixed effects. In contrast to existing SAE-methods, our proposed approach assists SAE-practitioners by an automated model selection. We highlight strengths and weaknesses of random forests in the context of SAE, in comparison to existing (or `traditional') SAE-methods using design- and model-based simulations. A distinct merit of this paper is the provision of a reliable bootstrap-scheme determining the uncertainty of area-level mean-estimates. Thus, this paper aims to contribute towards the trend of diversifying the model-toolbox for SAE-practitioners and researchers, while simultaneously respecting the methodological and structural nature of SAE.}

The general idea of tree-based methods in SAE is not entirely new. \citet{Anderson_etal2014} use district-level data from Peru to juxtapose the performance of LMM-based and tree-based methods for estimating population densities. \citet{Bilton_etal2017} use classification trees for categorical variables to incorporate auxiliary information from administrative data to survey data on household-poverty in Nepal. For a continuous variable \citet{DeMolinerGoga2018} estimate mean electricity consumption curves for sub-populations of households in France by using methods of LMMs and regression-based trees. \citet{MoConvilleetal2019} propose a regression-tree estimator for finite‐population totals, which can be viewed as an automatically-selected post‐stratification estimator. \citet{Dagdougetal2020} analyse theoretical properties of random forest in the context of complex survey data. \citet{Mendez2008} provides theoretical and empirical considerations for using random forests in the context of SAE and compares their performance with ‘traditional’ unit-level LMMs for the estimation of area-level means. Although we share the general idea of \citet{Mendez2008}, the approach of this paper differs in several ways: first of all, we leave the random forests algorithm \citep{Breiman2001} unchanged and explicitly estimate random effects to account for the hierarchical structure of the data. Secondly, the proposed framework of this paper is more flexible and potentially extendable to model more complex hierarchical dependency structures as well as spatial and temporal correlations. Additionally, the extension to other methods of machine learning is possible, such as support vector machines or gradient boosted trees.

The paper is organized as follows: Section \ref{sec:2} provides a methodological introduction to random forests and introduces MERFs based on \citet{Hajjem2014} as a method that effectively amalgamates random forests and the possibility to account for hierarchical dependencies of unit-level observations. Additionally, we motivate a general unit-level mixed model, treating LMMs in SAE as special cases. In Section \ref{sec:2.3}, we discuss the construction of area-level mean-estimates. Random forests promote the flexibility of predictive models, but their lack of distributional assumptions complicates inferences. As a result, Section \ref{sec:3} proposes a non-parametric bootstrap-scheme for the estimation of the area-level MSE. In Section \ref{sec:4}, we use model-based simulations under complex settings to extensively discuss and compare the performance of the proposed method for point- and MSE-estimates. We claim MERFs to be a valid alternative to existing methods for the estimation of SAE-means. In Section \ref{sec:5}, we use household income data of the Mexican state Nuevo León to estimate area-level averages and corresponding uncertainty estimates. We highlight modelling and robustness properties of our proposed methods. Section \ref{sec:5.3} proceeds with a design-based simulation, which asses the quality of results in the application of Section \ref{sec:5.2}. Furthermore, the design-based simulation contributes to a genuine demonstration of properties and advantages of MERFs in the context of SAE. Section \ref{sec:6} concludes and motivates further research.

\section{Theory and method}\label{sec:2}
In this section we propose a flexible, data-driven approach using random forests for the estimation of area-level means in the presence of unit-level survey data. The method requires a joint understanding of tree-based modelling techniques and concepts of SAE. We review the basic theory of random forest and discuss modifications to ensure their applicability to hierarchical data and subsequently to applications of SAE.

\subsection{Review of random forests}\label{sec:2.1}
Random forests combine individual decision trees \citep{Breiman_etal1984} to improve their joint predictive power, while simultaneously reducing their prediction variance \citep{Biau_Scornet2016, Breiman2001}. \citet{Breiman2001} extends his idea of Bagging \citep{Breiman1996bagging} - which combines predictions from single trees through a bootstrap and aggregation procedure - to random forests that apply bootstrap aggregation on decorrelated trees. Note that the two tuning parameters of random forests are the number of trees (controlling the number of bootstrap replications) and the number of variables to be selected as candidates for each split (controlling the degree of decorrelation). Because the forest is a combination of decorrelated trees, each aiming to minimize the prediction MSE, the optimal estimator for the random forest regression function $f()$ also minimizes the point-wise MSE. The minimizer under squared error loss in the regression context, is given by the conditional mean of target variable $y$ given the data \citep{Efron_Hastie2016, Wager_Athey2018}.

Random forests captivate with a lack of assumptions such as linearity or the distributional specification of model errors, however, observations are assumed to be independent. Applications of SAE are characterized by the use of hierarchical data. Ignoring the correlation between observations, generally results in inferior point-predictions and inferences. LMMs capture the dependencies between observations by random effects, while effects between covariates are modelled by linear fixed effects, resulting in an additive model of both terms. In the context of tree-based methods, \citet{Sela_Simonoff2012} propose a semi-parametric mixed model consisting of a random effects part and a fixed effects non-parametric tree-model. \citet{Hajjem_etal2011} propose a similar approach under the label of mixed effect regression trees (MERT). As the superior performance of random forests over regression trees transfers to dependent data, \citet{Hajjem2014} replace the fixed effects part in MERTs by a random forest, leading to mixed effects random forests (MERF). We scrutinize this approach and propose a general semi-parametric unit-level mixed model combining the flexibility of tree-based models with the structural advantages of linear mixed models in the next subsection.

\subsection{Mixed effects random forests}\label{sec:2.2}
We assume a finite population which is divided into $D$ disjunct areas $U_i$, with population sizes $N_i$, where $i = 1,...,D$ specifies the areas and $N = \sum_{i=1}^{D} N_i$ defines the population size. We assume to have a sample from this population. The number of sampled observations in area $i$ is given by $n_i$, where the sample size is denoted by $n = \sum_{i=1}^{D} n_i$. In each sampled area we obtain $j$ individual observations ranging from $1,...,n_i$.

We define the metric target variable for area $i$ as a $n_i \times 1$ vector of individual observations $y_i = [y_{i1},..., y_{in_i}]'$. Covariates are captured in the $n_i \times p$ matrix of $X_i = [x_{i1},..., x_{in_i}]'$, where $p$ defines the number of covariates.  $Z_i = [z_{i1},...,z_{in_i}]'$ defines the $n_i \times q$ matrix of area-specific random effect specifiers, where $q$ describes the dimension of random effects. $v_i = [v_{i1},...,v_{iq}]'$ is the $q \times 1$ vector of random effects for area $i$. $\epsilon_i = [\epsilon_{i1},...,\epsilon_{in_i}]' $ is the $n_i \times 1$ vector of individual error terms. Observations between areas are assumed to be independent and $v_i$ and $\epsilon_i$ are mutually independently normally distributed with the same variance-covariance matrix $H_i$ for random effects of each area $i$ and $R_i$ for individual errors. A joint notation for all $D$ areas is as follows:

\begin{align*}
	y = col_{1\leq i\leq D}(y_i) = (y_i',...,y_D')', \quad X = col_{1\leq i\leq D}(X_i), \\
	Z = diag_{1\leq i\leq D}(Z_i), \quad v = col_{1\leq i\leq D}(v_i), \quad \epsilon = col_{1\leq i\leq D}(\epsilon_i), \\
	R = diag_{1\leq i\leq D}(R_i), \quad H = diag_{1\leq i\leq D}(H_i).
\end{align*}

The goal is to identify a relation $f()$ between covariates $X$ and the target variable $y$, in order to predict values for non-sampled observations utilizing available supplementary covariates from census or register information across areas. We state a model consisting of two major parts: a fixed part $f(X)$ and a linear part $Zv$ capturing dependencies by random effects. In the following, $f()$ can be any parametric or non-parametric function that expresses the conditional mean of target variable $y$ given covariates $X$:

\begin{equation}\label{mod1}
	y = f(X)+Z v + \epsilon,
\end{equation}
where
$$ \epsilon \sim N(0,R) \quad \text{and}  \quad v \sim N(0,H). $$
Note that for each area $i$ the following model holds:

\begin{equation}
	y_i = f(X_i)+Z_i v_i + \epsilon_i.
\end{equation}

The covariance matrix of the observations $y$ is given by the block diagonal matrix $Cov(y) = V = diag_{1\leq i\leq D}(V_i)$, where $V_i = Z_i H_i Z_i' +R_i$. We introduce model (\ref{mod1}) in general terms to potentially allow for modelling of complex covariance and dependency structures. However, for the rest of the paper we assume that correlations arises only due to between-area variation, i.e. $R_i = \sigma_{\epsilon}^2 I_{n_i}$ for all areas $i$. Note that the already mentioned LMM proposed by \cite{Battese_etal1988} for estimating area-level means results as a special case of (\ref{mod1}) by setting $f()$ to be the linear model $f(X) = X\beta$, with regression parameters $\beta = [\beta_1,...,\beta_p]'$. Defining $f()$ as a random forest, results in the MERF-approach proposed by \citet{Hajjem2014}, which is the preferred specification throughout the rest of the paper.

Before we continue, we want to clarify consequences of distributional assumptions in (\ref{mod1}) that mainly address the linear part of the model. The unit-level errors are assumed to follow $\epsilon \sim N(0,R)$. However, the assumption of normality does not affect the properties of the fixed part $f(X)$ and we do not require residuals to be normally distributed for the application of our proposed method. Nevertheless, for the random components part, we require a proper likelihood function to ensure that the adapted expectation-maximization (EM) algorithm (see below for further details) for parameter estimates converges towards a local maximum within the parameter space. A normality-based likelihood function is exploited, as it has two important properties: firstly, it facilitates the estimation of random effects due to the existence of a closed-form solution of the integral of the Gaussian likelihood function. Secondly, the maximum likelihood estimate for the variance of unit-level errors is given by the mean of the unit-level residual sum of squares. The estimation of the random effects could be also done in a non-parametric way by using discrete mixtures \citep{Marino2016, Marino2019}. However, the modification towards a fully non-parametric formulation of model (\ref{mod1}) is subject to further research.

For fitting the model (\ref{mod1}) we use an approach reminiscent of the EM-algorithm similar to \cite{Hajjem2014}. In short, the MERF-algorithm subsequently estimates a) the forest function, assuming the random effects term to be correct and b) estimates the random effects part, assuming the Out-of-Bag-predictions (OOB-predictions) from the forest to be correct. OOB-predictions utilize the unused observations from the construction of each forest's sub-tree \citep{Breiman2001, Biau_Scornet2016}. The proposed algorithm is as follows:

\begin{enumerate}
	\item Initialize $b = 0$ and set random components $\hat{v}_{(0)}$ to zero.
	\item Set $b = b+1$. Update $\hat{f}(X)_{(b)}$ and $\hat{v}_{(b)}$:
	\begin{enumerate}
		\item $y^*_{(b)} = y -Z \hat{v}_{(b-1)}$
		\item Estimate $\hat{f}()_{(b)}$ using a random forest with dependent variable $y^*_{(b)}$ and covariates $X$. Note that $\hat{f}()_{(b)}$ is the same function for all areas $i$.
		\item Get the OOB-predictions $\hat{f}(X)^{OOB}_{(b)}$.
		\item Fit a linear mixed model without intercept and restricted regression coefficient of 1 for $\hat{f}(X)^{OOB}_{(b)}$:
		$$y = \hat{f}(X)^{OOB}_{(b)} +Z \hat{v}_{(b)} + \epsilon.$$
		\item Extract the variance components $\hat{\sigma}^2_{\epsilon,(b)}$ and $\hat{H}_{(b)}$ and estimate the random effects by:
$$\hat{v}_{(b)} = \hat{H}_{(b)}Z ' \hat{V}_{(b)}^{-1} (y - \hat{f}(X)^{OOB}_{(b)}).$$
	\end{enumerate}
	\item Repeat Step (2) until convergence is reached.
\end{enumerate}
The convergence of the algorithm is assessed by the marginal change of the modified generalized log-likelihood (GLL) criterion:
$$GLL (f,v_i | y) = \sum_{i=1}^{D}([y_i - f(X_i) - Z_i v_i ]' R_i^{-1} [ y_i - f(X_i) - Z_i v_i] + v_i ' H_i ^{-1} v_i +\text{log} |H_i|+ \text{log}|R_i|).$$

In the linear case with $f = X \beta$, and for given variance components $H$ and $R$, the maximization of the GLL-criterion is equivalent to the solution of so-called mixed model equations \citep{Wu_Zhang2006} - leading to best linear unbiased predictor (BLUP): $ \hat{v} = HZ' V ^{-1} (y - X\hat{\beta})$. For random forests, the corresponding estimator $\hat{v}$ for known parameters $H$ and $R$ is given by:
\begin{equation}\label{v_opt}
\hat{v} = HZ ' V ^{-1} (y - \hat{f}(X)^{OOB}).
\end{equation}
Mathematical details of the derivations are provided in Appendix A. This result is in line with \cite{capitaine_etal2021}, claiming that $\hat{v}$ is obtained by taking the conditional expectation given the data $y$ and subsequently $\hat{v}$ can thus be considered as the BLUP for the linear part of model (\ref{mod1}).

The estimation of variance components in Step 2 (d) for $\hat{\sigma}^2_{\epsilon}$ and $\hat{H}$ is obtained by taking the expectation of maximum likelihood estimators given the data. Although $\hat{\sigma}^2_{\epsilon}$ is a naive estimator within the discussed framework, it cannot be considered as a valid estimator for the variance $\sigma_{\epsilon}^2$ of the unit-level errors $\epsilon$. \citet{Breiman2001} maintains that the sum of squared residuals from OOB-predictions are a valid estimator for the squared prediction error of new individual observations. However, as an estimator of the residual variance under the model, $\hat{\sigma}^2_{\epsilon}$ is positively biased, as it includes uncertainty regarding the estimation of function $\hat{f}()$. Following \citet{Mendez_Lohr2011} we use a bias-adjusted estimator for the residual variance $\sigma^2_{\epsilon}$ from a random forest model (\ref{mod1}) using a bootstrap bias-correction. The essential steps to obtain the corrected residual variance are summarized as follows:

\begin{enumerate}
	\item Use the OOB-predictions $\hat{f}(X)^{\text{OOB}}$ from the final model $\hat{f}()$ after convergence of the algorithm.
	\item Generate $B$ bootstrap samples $y^{\star}_{(b)} = \hat{f}(X)^{\text{OOB}} + \epsilon^{\star}_{(b)}$, where the values $\epsilon^{\star}_{(b)}$ are sampled with replacement from the centred marginal residuals $\hat{e} = y -\hat{f}(X)^{\text{OOB}}$.
	\item Recompute $\hat{f}(X)^{\text{OOB}}_{(b)}$ using a random forest with $y^{\star}_{(b)}$ as dependent variable.
	\item Estimate the correction-term $K(\hat{f})$ by:
	\begin{align*}
		\hat{K}(\hat{f}) = B^{-1} \sum_{b=1}^{B} \left[\hat{f}(X)^{\text{OOB}} - \hat{f}(X)^{\text{OOB}}_{(b)}\right]^2.
	\end{align*}
\end{enumerate}

The bias-corrected estimator for the residual variance is then given by:
\begin{equation}\label{biasadj}
	\hat{\sigma}_{bc,\epsilon}^2 = \hat{\sigma}_{\epsilon}^2 - \hat{K}(\hat{f}).
\end{equation}

\subsection{Predicting small-area averages}\label{sec:2.3}
The MERF-model (\ref{mod1}) predicts the conditional mean on individual level of a metric dependent variable given unit-level auxiliary information. In the context of SAE, we are not interested in predictions on individual level, but in estimating indicators such as area-level means or area-level totals \citep{Rao_Molina2015}. Thus, we assume the same structural simplifications as the LMM proposed by \cite{Battese_etal1988} for estimating area-level means throughout the paper, i.e. $q=1$, $Z$ is a $n_i \times D$ design-matrix of area-intercept indicators, $v = [v_{1},...,v_{D}]'$ is a $D \times 1$ vector of random effects, and variance-covariance matrix for random effects simplifies to $H_i = \sigma_{v}^2$.

Firstly, we use the fact that random forest estimates of the fixed part $\hat{f}()$ express the conditional mean on unit-level. We calculate the mean-estimator for each area $i$ based on available supplementary data sources (usually census or administrative data) by: $$\bar{\hat{f}}(X_{i}) = \frac{1}{N_i} \sum_{j=1}^{N_i} \hat{f}(X_{i})=\frac{1}{N_i} \sum_{j=1}^{N_i} \hat{f}(x_{ij}).$$ Secondly, we exploit the result (\ref{v_opt}) that $\hat{v}_i$ is the BLUP for the linear part of the model (\ref{mod1}). Therefore, the proposed estimator for the area-level mean $\mu = [\mu_1,..., \mu_D]'$ is given by:

\begin{equation}\label{mu1}
	\hat{\mu}_i = \bar{\hat{f}}(X_i) +  Z_i\hat{v}_i\enspace\enspace\text{for}\enspace\enspace i=1,...D.
\end{equation}

In cases of non-sampled areas, the proposed estimator for the area-level mean reduces to the fixed part from the random forest:
$$\hat{\mu}_i = \bar{\hat{f}}(X_i).$$

\textcolor{black}{We shortly discuss properties of our proposed estimator from Equation (\ref{mu1}). The structural component $Z_i\hat{v}_i$ captures dependency and correlation structures by random effects and the expression $\bar{\hat{f}}(X_i)$ is the fixed effects predictor of the mean. For the special case, where $\hat{f}()$ is assumed to be the linear model $\hat{f}(X) = X\hat{\beta}$, with regression parameters $\hat{\beta} = [\hat{\beta_1},...,\hat{\beta}_p]'$, the estimator for area-level means resembles the result of the EBLUP \citep{Battese_etal1988}. If $\hat{f}()$ is a random forest, we face area-specific mean-estimates for fixed-effects from a highly flexible, data-driven and non-differentiable function. Two major tuning parameters affect the predictive performance of the random forest  $\hat{f}()$, i.e. the number of trees and the number of split-candidates at each node controlling the degree of decorrelation. In contrast to existing parametric and non-parametric methods in SAE, our proposed estimator from Equation (\ref{mu1}) abstains from problems due to model-selection. Random forests implicitly perform optimized model-selection including higher-order effects or non-linear interactions. Although flexible approaches such as P-Splines \citep{Opsomeretal2008} potentially determine non-linear relations in covariates, users have to explicitly specify model-variables and interactions to be interpolated a-priori, resulting in a comparable paradigm of model-selection compared to standard LMMs. An additional property of $\hat{f}()$ is the capability to deal with high-dimensional covariate data, i.e. cases where the number of covariates $p$ is larger than the sample size $n$. This property might be exploited in the context of applications to alternative Big Data sources \citep{Marchetti_etal2015,Schmid2017}.}

\section{Estimation of uncertainty}\label{sec:3}
The assessment of uncertainty of area-level indicators in SAE is crucial to analyse the quality of estimates. The area-level MSE is a conventional measure fulfilling this goal, but its calculation is a challenging task. For instance, for the unit-level LMM with block diagonal covariance matrices \citep{Battese_etal1988}, the exact MSE cannot be analytically derived with estimated variance components \citep{Gonzalez_etal2008, Rao_Molina2015} and only partly-analytical approximations are available \citep{Prasad_Rao1990,Datta_Lahiri2000}. An alternative to estimate uncertainty of the area-level indicators are bootstrap-schemes \citep{Hall_Maiti2006, Gonzalez_etal2008, Chambers_Chandra2013}. In contrast, general statistical results for inference of random forests are rare, especially in comparison to the existing theory of inference using LMMs. \textcolor{black}{Nevertheless, we provide a theoretical discussion on the estimation of MSEs for in-sample area-level means in the spirit of \citet{Prasad_Rao1990} based on \citet{Mendez2008}. Derivations can be found in the online supplementary materials. The resulting analytical approximation is considered to be a complement to contextualize the quality of our proposed bootstrap MSE-estimator for in-sample areas. We discuss performance details in the model-based simulation in Section \ref{sec:4}. An exact theoretical determination and discussion of asymptotic properties will be left to further research.} The theoretical background of random forests grows, but mainly aims to quantify the uncertainty of individual predictions \citep{Sexton_Laake2009, wager_etal2014, Wager_Athey2018, Athey_etal2019, Zhang2019}. The extension of recent theoretical results, such as conditions for the consistency of unit-level predictions \citep{Scornet_etal2015} or their asymptotic normality \citep{Wager_Athey2018}, towards area-level indicators is a conducive strategy.

In this paper, we propose a non-parametric random effect block (REB) bootstrap for estimating the MSE of the introduced area-level estimator given by Equation (\ref{mu1}). We aim to capture the dependence-structure of the data as well as the uncertainty introduced by the estimation of model (\ref{mod1}). Our bootstrap-scheme builds on the non-parametric bootstrap introduced by \cite{Chambers_Chandra2013}. The proposed REB bootstrap has two major advantages: firstly, empirical residuals depend only on the correct specification of the mean behaviour function $f()$ of the model, thus the REB setting is lenient to specification errors regarding the covariance structure of the model. Secondly, the bootstrap within blocks ensures that the variability of residuals within each area is captured. We scale and centre the empirical residuals by the bias-corrected residual variance (\ref{biasadj}) in order to eliminate the uncertainty due to the estimation of $\hat{f}()$ from the naive residuals. The steps of the proposed bootstrap are as follows:

\begin{enumerate}
	\item For given $\hat{f}()$ calculate the $n_i\times 1$ vector of marginal residuals $\hat{e}_i = y_i -\hat{f}(X_i)$ and define $\hat{e} = [\hat{e}_1',...,\hat{e}_D']'$.
	\item Using the marginal residuals $\hat{e}$, compute level-2 residuals for each area by $$\bar{r}_{i} = \frac{1}{n_i} \sum_{j=1}^{n_i} {\hat{e}_{i}}\enspace\enspace\text{for}\enspace\enspace i=1,...D$$ and $\bar{r} = [\bar{r}_1,...,\bar{r}_D]'$ indicates the $D\times 1$ vector of level-2 residuals.
\item To replicate the hierarchical structure we use the marginal residuals and obtain the $n_i\times 1$ vector of level-1 residuals by $\hat{r}_{i} = \hat{e}_{i} - 1_{n_i}\bar{r}_i$. The residuals $\hat{r} = [\hat{r}_1',...,\hat{r}_D']'$ are scaled to the bias-corrected variance $\hat{\sigma}_{bc,\epsilon}^2$ (\ref{biasadj}) and centred, denoted by $\hat{r}^{c} = [\hat{r}^{c'}_{1},...,\hat{r}^{c'}_{D}]'$. The level-2 residuals $\bar{r}_i$ are also scaled to the estimated variance $\hat{H}_i=\hat{\sigma}_{v}^2$ and centred, denoted by $\bar{r}^{c} = [\bar{r}^{c}_1,...,\bar{r}^{c}_D]'$.
	\item For $b=1,...,B$:
	\begin{enumerate}
		\item Sample independently with replacement from the scaled and centred level-1 and level-2 residuals:
\begin{eqnarray}
  \nonumber r_{i}^{(b)}=\text{srswr}(\hat{r}^c_{i},n_i)\enspace\enspace \text{and}\enspace\enspace \bar{r}^{(b)}=\text{srswr}(\bar{r}^c,D).
\end{eqnarray}
		\item Calculate the bootstrap population as $y^{(b)} = \hat{f}(X) +Z \bar{r}^{(b)}+ r^{(b)}$ and calculate the true bootstrap population area means $\mu_i^{(b)}$ as $\frac{1}{N_i} \sum_{j=1}^{N_i} y_{ij}^{(b)}$ for all $i = 1,..,D$.
		\item For each bootstrap population draw a bootstrap sample with the same $n_i$ as the original sample. Use the bootstrap sample to obtain estimates $\hat{f}^{(b)}()$ and $\hat{v}^{(b)}$ as discussed in Section \ref{sec:2.2}.
		\item Calculate area-level means following Section \ref{sec:2.3} by $$\hat{\mu}^{(b)}_{i} = \bar{\hat{f}}^{(b)}(X_i) +  Z_i\hat{v}_i^{(b)}.$$
	\end{enumerate}
	\item Using the $B$ bootstrap samples, the MSE-estimator is obtained as follows:
	$$\widehat{MSE}_i = B^{-1} \sum_{b=1}^{B} \left(\mu_i^{(b)}-\hat{\mu}^{(b)}_{i}\right)^2.$$
\end{enumerate}

\section{Model-based simulation}\label{sec:4}
This section marks the first step in the empirical assessment of the proposed method. The model-based simulation juxtaposes point estimates for the area-level mean from the mixed effects random forest model (\ref{mod1}) with several competitors. In particular, we study the performance of MERFs compared to the BHF-model \citep{Battese_etal1988}, the EBP \citep{Molina_rao2010}\textcolor{black}{, the EBP under data-driven Box-Cox transformation (EBP-BC) by \citet{Rojas_etal2019} as well as the non-parametric EBLUP with P-Splines (P-SPLINES) by \citet{Opsomeretal2008}}. The BHF-model serves as an established baseline for the estimation of area-level means and the EBP and the EBP-BC conceptually build on the BHF-model. \textcolor{black}{The non-parametric EBLUP by \citet{Opsomeretal2008} incorporates advantages of flexible, non-linear smoothing methods into existing theory for SAE based on LMMs.} Differences in the performance of the EBP and the EBP-BC highlight advantages of data-driven transformations, while differences in the performance of the linear competitors and \textcolor{black}{more flexible alternatives (MERF, P-SPLINES) indicate advantages of semi-parametric and non-linear modelling}. Overall, we aim to show, that our proposed methodology for point and uncertainty estimates performs comparably well to `traditional' SAE-methods and has comparative advantages in terms of robustness against model-failure.

The simulation-setting is characterized by a finite population $U$ of size $N=50000$ with $D=50$ disjunct areas $U_1,...,U_D$ of equal size $N_i = 1000$. We generate samples under stratified random sampling, utilizing the $50$ small areas as stratas, resulting in a sample size of $n = \sum_{i=1}^{D} n_i = 1229$. The area-specific sample sizes range from $6$ to $49$ sampled units with a median of $21$ and a mean of $25$. The sample sizes are comparable to area-level sample sizes in the application in Section \ref{sec:5}  and can thus be considered to be realistic.

We consider four scenarios denoted as \textit{Normal}, \textit{Interaction}, \textit{Normal-Par}, \textit{Interaction-Par} and repeat each scenario independently $M=500$ times. The comparison of competing model-estimates under these four scenarios allows us to examine the performance under two major dimensions of model-misspecification: Firstly, the presence of skewed data delineated by non-normal error-terms and secondly, the presence of unknown non-linear interactions between covariates.
Scenario \textit{Normal} provides a baseline under LMMs with normally distributed random effects and unit-level errors. As model-assumptions for LMMs are fully met, we aim to show, that MERFs perform comparably well to linear competitors in the reference scenario. Scenario \textit{Interaction} shares its error-structure with \textit{Normal}, but involves a complex model including quadratic terms and interactions. This scenario portrays advantages of semi-parametric and non-linear modelling methods protecting against model-failure. Working with inequality or income data, we often deal with skewed target variables. Thus, we use the Pareto distribution to mimic realistic income scenarios. Scenario \textit{Normal-Par} uses the linear additive structure of LMMs and adds Pareto distributed unit-level errors. The resulting scenario, including a skewed target variable, is a classical example promoting the use of transformations assuring that assumptions of LMMs to be met. Finally, scenario \textit{Interaction-Par} combines the two discussed dimensions of model misspecification, i.e. a non-Gaussian error-structure with complex interactions between covariates. We chose this scenario to emphasize the ability of MERFs to handle both complications simultaneously. Further details on the data-generating process for each scenario are provided in Table \ref{tab:MB1}.

\begin{table}[ht]
	\centering
	\captionsetup{justification=centering,margin=1.5cm}
   \caption{Model-based simulation scenarios}
	\resizebox{\textwidth}{!}{\begin{tabular}{rlccccccc}
			\toprule
			{Scenario} & {Model} & {$x1$} & {$x2$} & {$\mu$} & {$v$} & {$\epsilon$} \\ \midrule
			Normal  & $ y = 5000-500x_1-500x_2+v+\epsilon$ & $N(\mu,3^2)$ & $N(\mu,3^2)$ & $unif(-1,1)$  & $N(0,500^2)$ & $N(0,1000^2)$  \\
			Interaction  & $ y = 15000-500x_1x_2-250x_2^2+v+\epsilon $  & $N(\mu,4^2)$ & $N(\mu,2^2)$  & $unif(-1,1)$ & $N(0,500^2)$ &$N(0,1000^2)$ \\
			Normal-Par  & $ y = 5000-500x_1-500x_2+v+\epsilon $  & $N(\mu,3^2)$ & $N(\mu,3^2)$  & $unif(-1,1)$ & $N(0,500^2)$ & $Par(3,800)$ \\
			Interaction-Par  & $ y = 20000 -  500x_1x_2 - 250x_2^2+ v + \epsilon $  & $N(\mu,2^2)$ & $N(\mu,2^2)$  & $unif(-1,1)$ & $N(0,1000^2)$  & $Par(3,800)$ \\ \bottomrule
	\end{tabular}}
\label{tab:MB1}
\end{table}

We evaluate point estimates for the area-level mean by the relative bias (RB) and the relative root mean squared error (RRMSE). As quality-criteria for the evaluation of the MSE-estimates, we choose the relative bias of RMSE (RB-RMSE) and the relative root mean squared error of the RMSE:
\begin{align}
\nonumber	RB_i &= \frac{1}{M} \sum_{m=1}^{M} \left(\frac{\hat{\mu}^{(m)}_i - \mu^{(m)}_i}{\mu^{(m)}_i}\right)\\\nonumber
\textcolor{black}{RRMSE_i} &= \textcolor{black}{\frac{\sqrt{\frac{1}{M} \sum_{m=1}^{M} \left(\hat{\mu}^{(m)}_i - \mu^{(m)}_i\right)^2}}{\frac{1}{M}\sum_{m=1}^{M}\mu^{(m)}_i}}\\\nonumber
RB\text{-}RMSE_i &=\frac{\sqrt{\frac{1}{M} \sum_{m=1}^{M} MSE^{(m)}_{est_i}} - RMSE_{emp_i}}{RMSE_{emp_i}}\\\nonumber
RRMSE\text{-}RMSE_i &= \frac{\sqrt{\frac{1}{M} \sum_{m=1}^{M} \left(\sqrt{MSE^{(m)}_{est_i}} - RMSE_{emp_i}\right)^2}}{RMSE_{emp_i}},
\end{align}

where $\hat{\mu}^{(m)}_i$ is the estimated mean in area $i$ based on any of the methods mentioned above and $\mu^{(m)}_i$ defines the true mean for area $i$ in simulation round $m$. $MSE_{est_i}^{(m)}$ is estimated by the proposed bootstrap in Section \ref{sec:3} and $RMSE_{emp_i} = \sqrt{\frac{1}{M} \sum_{m=1}^{M}(\hat{\mu}^{(m)}_i -\mu^{(m)}_i)^2}$ is the empirical root MSE over $M$ replications.

For the computational realization of the model-based simulation, we use R \citep{R_language}. The BHF estimates are realized from the \emph{sae}-package \citep{Molina_Marhuenda:2015} and the \emph{emdi}-package \citep{Kreutzmann_etal2019} is used for the EBP as well as the EBP under the data-driven Box-Cox transformation. \textcolor{black}{We implement the P-SPLINE method with the package \emph{mgcv} \citep{Wood_2017}.} For estimating the proposed MERF-approach, we use the packages \emph{randomForest} \citep{Liaw_Wiener2002} and \emph{lme4} \citep{Bates_etal2015}. We monitor the convergence of algorithm introduced in Section \ref{sec:2.2} with a precision of $1e^{-5}$ in relative difference of the GLL-criterion and set the number of split-candidates to $1$, keeping the default of $500$ trees for each forest.

\begin{figure}[ht]
	\centering
	\captionsetup{justification=centering,margin=1.5cm}
		\includegraphics[width=1\linewidth]{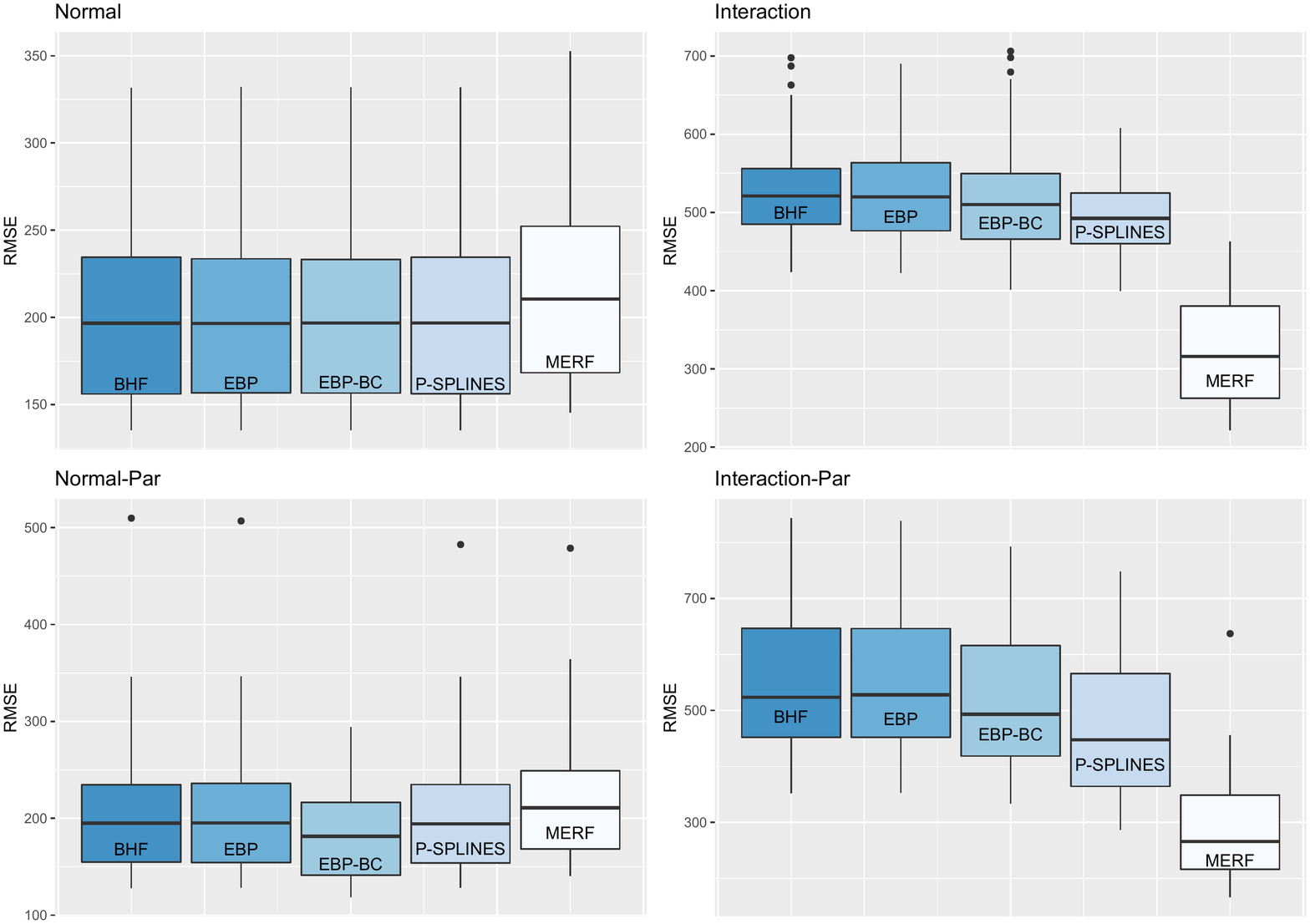}
	\caption{Empirical RMSE comparison of point estimates for area-level averages under four scenarios}
	\label{fig:MBpoint}
\end{figure}

We start with a focus on the performance of point estimates. Figure \ref{fig:MBpoint} reports the empirical RMSE of each method under the four scenarios. As expected, in the \textit{Normal} scenario, the BHF and the EBP perform on the same level and outperform the MERF estimator. The EBP with a data-driven transformation (EBP-BC) \textcolor{black}{and the non-parametric EBLUP (P-SPLINES)} lead to similar results compared to the BHF and EBP. This shows that the data-driven transformation \textcolor{black}{and the penalized smoothing approach} work as expected. A similar pattern appears in the results from the \textit{Normal-Par} scenario, except that the EBP-BC reaches a lower overall RMSE due to its property of data-driven transformation and subsequently improved estimation under skewed data. As anticipated, a comparison of the performance of the MERF between the \textit{Normal} and the \textit{Normal-Par} scenario indicates, that the MERF exhibits robust behaviour under skewed data and subsequently regarding violations of the normal distribution of errors. \textcolor{black}{LMM-based competitors match the data-generating process of fixed effects and perform accordingly, as already observed under the \textit{Normal} scenario.}
For complex scenarios, i.e. \textit{Interaction} and \textit{Interaction-Par}, point estimates of the proposed MERF outperform the SAE-methods based on LMMs. The EBP-BC performs better in terms of lower RMSE values compared to the BHF and the EBP in both interaction scenarios. \textcolor{black}{The flexible approach of P-SPLINES outperforms the BHF, the EBP and the data-driven EBP-BC. However, MERFs automatically identify interactions and non-linear relations, such as the quadratic term in scenario \textit{Interaction-Par}, which leads to a clear comparative advantage in terms of RMSE.} Overall, the results from Figure \ref{fig:MBpoint} indicate that the MERF performs comparably well to LMMs in simple scenarios, and outperforms `traditional' SAE-models in the presence of unknown non-linear relations between covariates. Additionally, the robustness against model-misspecification of MERFs holds if distributional assumptions for LMMs are not met, i.e. in the presence of non-normally distributed errors and skewed data. Table \ref{tab:MBpoint} reports the corresponding values of RB and RRMSE for our discussed point estimates. The RB and the RRMSE from the MERF-method attest a competitively low level for all scenarios. Most interestingly, in complex scenarios (\textit{Interaction} and \textit{Interaction-Par}), a familiar result regarding the statistical properties of random forests appears: the RB is higher compared to the LMM-based models, but the enlarged RB is rewarded by a lower RRMSE for point estimates.

\begin{table}[!h]
\footnotesize
	\centering
	\captionsetup{justification=centering,margin=1.5cm}
	\caption{Mean and Median of RB and RRMSE over areas for point estimates in four scenarios}
	\begin{tabular}{@{\extracolsep{5pt}} lrcccccccc}
			\\[-1.8ex]\hline
			\hline \\[-1.8ex]
			& &\multicolumn{2}{c}{\textit{Normal}} &\multicolumn{2}{c}{\textit{Interaction}}&\multicolumn{2}{c}{\textit{Normal-Par}}&\multicolumn{2}{c}{\textit{Interaction-Par}} \\
			\hline \\[-1.8ex]
			& & Median & Mean & Median & Mean & Median & Mean & Median & Mean \\
			\hline \\[-1.8ex]
			\multicolumn{9}{l}{RB[\%]}\\
			\hline \\[-1.8ex]
			&BHF & $0.087$ & $0.131$ & $$-$0.202$ & $0.106$ & $0.193$ & $0.220$ & $0.043$ & $0.142$ \\
			&EBP & $0.069$ & $0.128$ & $$-$0.060$ & $0.108$ & $0.216$ & $0.217$ & $0.105$ & $0.142$ \\
			&EBP-BC & $0.152$ & $0.184$ & $0.156$ & $0.381$ & $0.174$ & $0.129$ & $0.139$ & $0.262$ \\
			&P-SPLINES & $0.096$ & $0.137$ & $$-$0.064$ & $0.123$ & $0.199$ & $0.227$ & $0.051$ & $0.090$ \\
			&MERF & $0.137$ & $0.191$ & $0.279$ & $0.312$ & $0.409$ & $0.460$ & $0.151$ & $0.188$ \\
		 	\hline \\[-1.8ex]
		 	\multicolumn{9}{l}{RRMSE[\%]}\\
		 	\hline \\[-1.8ex]			
        &BHF & $3.830$ & $4.090$ & $3.770$ & $3.870$ & $3.600$ & $4.100$ & $2.800$ & $2.950$ \\
		&EBP & $3.850$ & $4.100$ & $3.750$ & $3.870$ & $3.600$ & $4.120$ & $2.830$ & $2.950$ \\
		&EBP-BC & $3.850$ & $4.100$ & $3.680$ & $3.800$ & $3.430$ & $3.710$ & $2.650$ & $2.770$ \\
		&P-SPLINES & $3.840$ & $4.090$ & $3.580$ & $3.620$ & $3.590$ & $4.100$ & $2.380$ & $2.490$ \\
		&MERF & $4.070$ & $4.380$ & $2.270$ & $2.330$ & $3.890$ & $4.380$ & $1.420$ & $1.530$ \\
			\hline \\[-1.8ex]
	\end{tabular}
\label{tab:MBpoint}
\end{table}

We scrutinize the performance of our proposed MSE-estimator on the four scenarios, examining whether the observed robustness against model-misspecification due to unknown complex interactions between covariates or skewed data for point estimates, also holds for our non-parametric bootstrap-scheme. For each scenario and each simulation round, we choose the parameter of bootstrap replications $B = 200$. From the comparison of RB-RMSE among the four scenarios provided in Table \ref{tab:MBmse}, we infer, that the proposed non-parametric bootstrap procedure effectively handles scenarios that lead to model-misspecification in cases of (untransformed) LMMs. This is demonstrated by essentially unbiasedness in terms of mean and median values of RB-RMSE over areas of the MSE-estimator under all four scenarios: independently, whether the data generating process is characterized by complex interactions (\textit{Interaction}), non-normal error terms (\textit{Normal-Par}) or a combination of both problems (\textit{Interaction-Par}). \textcolor{black}{We compare the performance of our bootstrap estimator to an estimator resulting from an analytical discussion of uncertainty in the spirit of \citet{Prasad_Rao1990}, which can be found in the online supplementary materials. The analytical approximation generally underestimates the MSE, except for the \textit{Interaction} scenario, which substantiates the quality of the proposed bootstrap estimator. 
A detailed graphical comparison of the RB-RMSE between the non-parametric bootstrap and the analytical MSE estimator is provided by Figure \ref{fig:MBappendix} in the Appendix B.}

\begin{table}[!h]
	\footnotesize
	\centering
	\captionsetup{justification=centering,margin=1.5cm}
	\caption{Performance of bootstrap and analytical MSE estimators in model-based simulation: mean and median of RB-RMSE and RRMSE-RMSE over areas}
	\begin{tabular}{@{\extracolsep{5pt}} lrcccccccc}
			\\[-1.8ex]\hline
			\hline \\[-1.8ex]
			& &\multicolumn{2}{c}{\textit{Normal}} &\multicolumn{2}{c}{\textit{Interaction}}&\multicolumn{2}{c}{\textit{Normal-Par}}&\multicolumn{2}{c}{\textit{Interaction-Par}} \\
			\hline \\[-1.8ex]
			& & Median & Mean & Median & Mean & Median & Mean & Median & Mean \\
			\hline \\[-1.8ex]
			\multicolumn{9}{l}{RB-RMSE[\%]}\\
			\hline \\[-1.8ex]
			&Bootstrap & $0.319$ & $$-$0.084$ & $0.127$ & $0.548$ & $0.340$ & $0.724$ & $$-$0.802$ & $0.123$ \\
			&Analytic & $$-$5.700$ & $$-$5.010$ & $0.707$ & $0.261$ & $$-$4.020$ & $$-$4.480$ & $$-$7.500$ & $$-$7.000$ \\
			\hline \\[-1.8ex]
			\multicolumn{9}{l}{RRMSE-RMSE[\%]}\\
			\hline \\[-1.8ex]			
			&Bootstrap & $12.500$ & $12.500$ & $22.200$ & $22.800$ & $43.100$ & $48.200$ & $41.000$ & $44.700$ \\
			&Analytic & $6.130$ & $5.930$ & $10.400$ & $12.200$ & $21.300$ & $21.400$ & $33.600$ & $33.500$ \\
			\hline \\[-1.8ex]
		\end{tabular}
\label{tab:MBmse}
\end{table}

\begin{figure}[ht]
	\centering
	\captionsetup{justification=centering,margin=1.5cm}
	\includegraphics[width=1\linewidth]{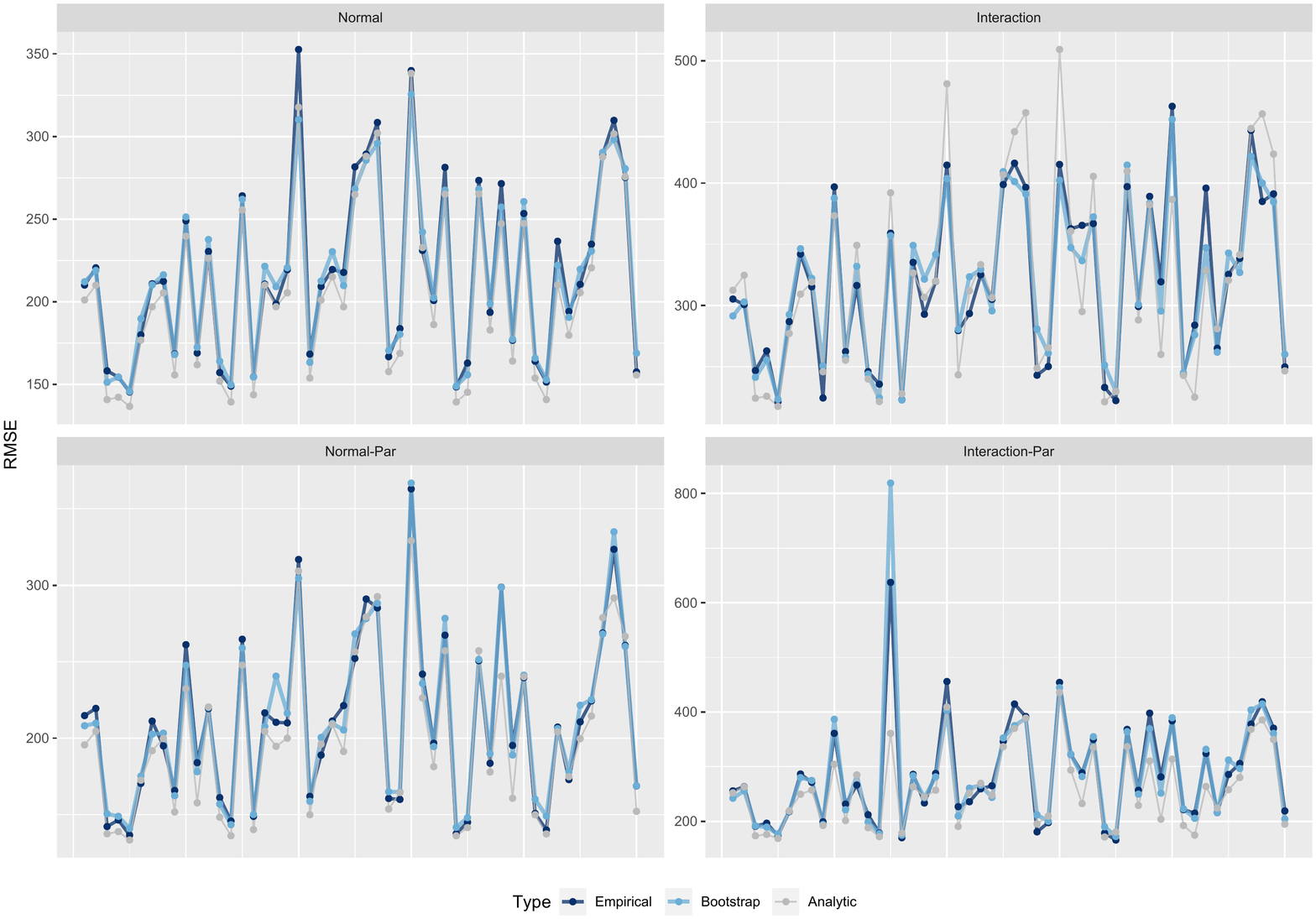}
	\caption{Empirical, bootstrapped and analytical area-level RMSEs for four scenarios}
	\label{fig:trackMSE}
\end{figure}

From the results of Table \ref{tab:MBmse} and the subsequent discussion, we cannot directly infer the area-wise tracking properties of the estimated RMSE against the empirical RMSE over our $500$ simulation rounds. Thus, Figure \ref{fig:trackMSE} provides additional intuition on the quality of our proposed non-parametric MSE-bootstrap estimator. Given the tracking properties in all four scenarios, we conclude that our bootstrap MSE-estimates strongly correspond to the empirical RMSE \textcolor{black}{and appear to track the domain-specific empirical RMSE more precisely than the estimates of our analytical MSE estimator from the theoretical discussion in the online supplementary materials.} Furthermore, we do not observe systematic differences between the bootstrapped and empirical MSE-estimates regarding different survey-sample sizes.

\section{Application: Estimating household income for Mexican municipalities}\label{sec:5}

In this section, we discuss the performance of our proposed method in the context of a genuine SAE example. Concretely, we apply the MERF-method proposed in Section \ref{sec:2.2} to estimate domain-level average household income for the Mexican state Nuevo León. Section \ref{sec:5.1} describes the data and Section \ref{sec:5.2} reports results. We end our empirical analysis by an additional design-based simulation enabling a profound discussion on the quality of point and MSE-estimates in Section \ref{sec:5.3}.

\subsection{Data description}\label{sec:5.1}
Income inequality in Mexico is a research topic of timeless importance, particularly regarding the effectiveness of developmental and social policies \citep{lambert_hyunmin2019}. Although the level of income inequality in Mexico is comparable to other Latin American countries, it is one of the highest among other OECD countries \citep{Oecd_21}. Analysing average national income values is a common practice, but constitutes an inappropriate measure to monitor the efficacy of regional policy measures. Besides detailed disaggregated information, also suitable statistical methods are needed to quantify local developments. For the following application, we break down regional differences in average household per capita income of one of 32 Mexican states. Nuevo León  is located in the North-East of Mexico and according to the (sub-national) Human Development Index (HDI), it is one of the most developed states of Mexico \citep{smits_Permanyer2019}. Nevertheless, the distribution of individual household income in Nuevo León is unequal and thus highly skewed. For instance, the Gini-coefficient of household income is comparable to the total Gini of Mexican household disposable income from 2012 which was $0.46$ \citep{Oecd_21}.

We use data from 2010 provided by CONEVAL (Consejo Nacional de Evaluación de la Política de Desarrollo Social), combining the Mexican household income and expenditure survey (Encuesta Nacional de Ingreso y Gastos de los Hogares, ENIGH) with a sample of census microdata by the National Institute of Statistics and Geography (Instituto Nacional de Estadística y Geografía). The dataset comprises income and socio-demographic data, equally measured by variables that are part of the survey as well as the census data. The target variable for the estimation of domain-level average household income in Section \ref{sec:5.2}, is the total household per capita income (\textit{ictpc}, measured in pesos), which is available in the survey but not in the census.

Nuevo León is divided into $51$ municipalities. While the census dataset in our example comprises information on $54848$ households from all $51$ municipalities, the survey data includes information on $1435$ households from $21$ municipalities, ranging from a minimum of $5$ to a maximum of $342$ households with a median of $27$ households. This leaves $30$ administrative divisions to be out-of-sample. Table \ref{tab:Apdetails} provides details on sample and census data properties.

\begin{table}[ht]
	\centering
	\captionsetup{justification=centering,margin=1.5cm}
	\caption{Summary statistics on in- and out-of-sample areas: area-specific sample size of census and survey data}
	\begin{tabular}{@{\extracolsep{5pt}} lcccccccc}
		\\[-1.8ex]\hline
		\hline \\[-1.8ex]
		&\multicolumn{2}{c}{Total}&\multicolumn{2}{c}{In-sample}&\multicolumn{2}{c}{Out-of-sample}\\
		&\multicolumn{2}{c}{51} & \multicolumn{2}{c}{21} & \multicolumn{2}{c}{30} \\ \hline
		\hline \\[-1.8ex]
		& Min. & 1st Qu. & Median & Mean & 3rd Qu. & Max. \\
		\hline
		Survey area sizes & 5.00 & 14.00 & 27.00 & 68.33 & 79.00 & 342.00 \\
		Census area sizes & 76.00 & 454.50 & 642.00 & 1075.45 & 872.50 & 5904.00 \\
		\hline
	\end{tabular}
\label{tab:Apdetails}
\end{table}

With respect to the design-based simulation in Section \ref{sec:5.3}, we emphasize that we are in the fortunate position of having a variable that is highly correlated with the target variable \textit{ictpc} in the application and that is available in the survey and in the census dataset: the variable \textit{inglabpc} measures earned per capita income from work. Although \textit{inglabpc} deviates from the desired income definition for our approach - as it covers only one aspect of total household income - it is effective to evaluate our method under a design-based simulation in Section \ref{sec:5.3}. Furthermore, the design-based simulation implicitly assess the quality of our empirical results from Section \ref{sec:5.2}.

Using data from Nuevo León for the estimation of domain-level income averages, is an illustrative and realistic example and imposes several challenges on the proposed method of MERFs: first of all, about $24$ percent of households in the survey-sample are located in the capital Monterrey. Secondly, there exist more out-of sample domains than in-sample domains. Moreover, we are confronted with households reporting zero-incomes. Our intention in choosing this challenging example for the application part in Section \ref{sec:5.2} and the following design-based simulation in Section \ref{sec:5.3} are simple: we aim to show, that our proposed approaches for point and uncertainty estimates demonstrate a valid alternative to existing SAE-methods and are predominantly applicable for cases where `traditional' methods perform poorly or even fail. Additionally, we aim to provide a clear-cut presentation and empirical assessment of MERFs for SAE, which requires a transparent discussion of advantages and potential weaknesses in demanding real-world examples. \\

\subsection{Results and discussion}	\label{sec:5.2}
\begin{figure}[ht]
	\centering
	\captionsetup{justification=centering,margin=1.5cm}
	\includegraphics[width=1\linewidth]{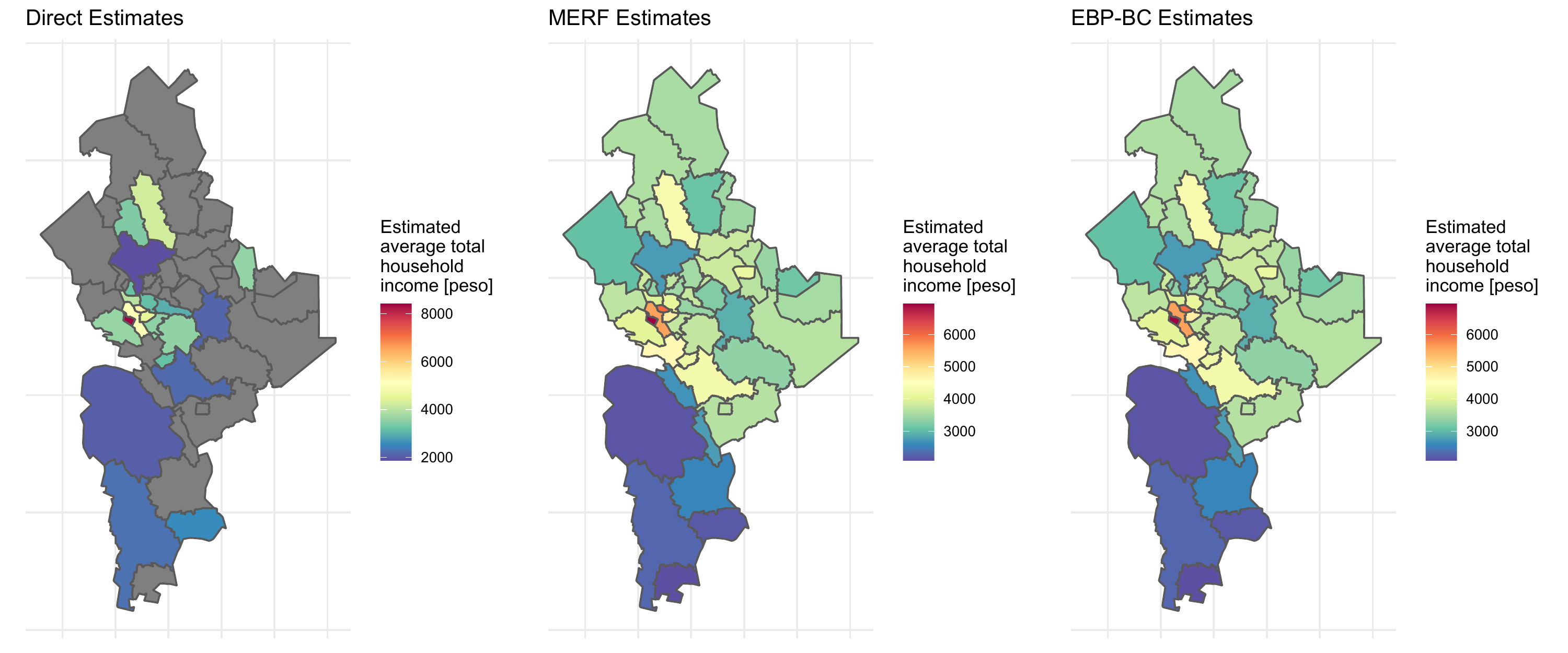}
	\caption{Estimated average total household per capita income \textit{itcpc} for the state Nuevo León based on three different estimation methods}
	\label{fig:mapincome}
\end{figure}

Direct estimates for average total household per capita income for Nuevo León are possible for 21 out of 51 domains. The use of model-based SAE-methods incorporating covariate census data will not only lead to estimates for the remaining out-of-sample areas, but correspondingly improve the overall quality of estimates \citep{Tzavidis_etal2018}. As variable \textit{ictpc} is highly skewed, we deduce potential issues of model-misspecification and suggest the use of the EBP-BC and the MERF. Given the theoretical discussion and the results in the model-based simulation in Section \ref{sec:4}, we infer that the EBP-BC and the proposed method of MERFs for SAE effectively handle non-normally distributed data. Moreover, we are particularly interested in differences between these two diverse SAE-models in the context of real-world applications. \textcolor{black}{We use the design-based simulation in Section \ref{sec:5.3} to extend our methodological discussion towards all methods discussed in the model-based simulation in Section \ref{sec:4}.} Figure \ref{fig:mapincome} maps results from direct estimates, the MERF and the EBP-BC. Obviously, the model-based estimates from the MERF and the EBP-BC expand the perspective of regional disparities in average total household income per capita towards non-sampled regions. Furthermore, we identify three distinct clusters of income levels from our results in Figure \ref{fig:mapincome} that had not been observable from the mapping of direct estimates: a low income cluster in the South of Nuevo León, a very high income cluster in the metropolitan area of the capital Monterrey and a group of middle-income areas between the North and the South of the state. This finding illustrates, the potential of model-based techniques to highlight patterns of regional income disparities and enable the mapping of reliable empirical evidence. Given the information provided by the three maps, we do not report major differences between the point estimates of MERFs and the EBP-BC.

Apart from mapping empirical results of domain averages, we are mainly interested in quality criteria, such as the coefficients of variation (CV) and the details of model-specification for the EBP-BC and the MERF. To obtain estimates of variances for the direct estimates, we use the calibrated bootstrap method \citep{Alfons_Templ2013} as provided in the R package \emph{emdi} \citep{Kreutzmann_etal2019}. \textcolor{black}{For the MSE-estimates of the MERF, we rely on the non-parametric bootstrap from Section \ref{sec:3}}. For the model of the data-driven EBP-BC, we follow the approach of \citet{Rojas_etal2019} and use the Bayesian Information Criterion (BIC) to identify valid predictors for the target variable of \textit{ictpc}. The resulting working-model includes variables determining occupation, sources of income, the socio-economic level and educational aspects of individual households. The identification of predictive covariates for MERFs highlights a conceptual difference to LMM-based methods. Due to the properties of the random forest algorithm \citep{Breiman2001}, random forests perform an implicit variable selection. The selected model for fixed effects in our case is characterized by an R-squared of about $0.47$ percent.

\begin{figure}[!htb]
	\centering
	\captionsetup{justification=centering,margin=1.5cm}
	\includegraphics[width=0.93\linewidth]{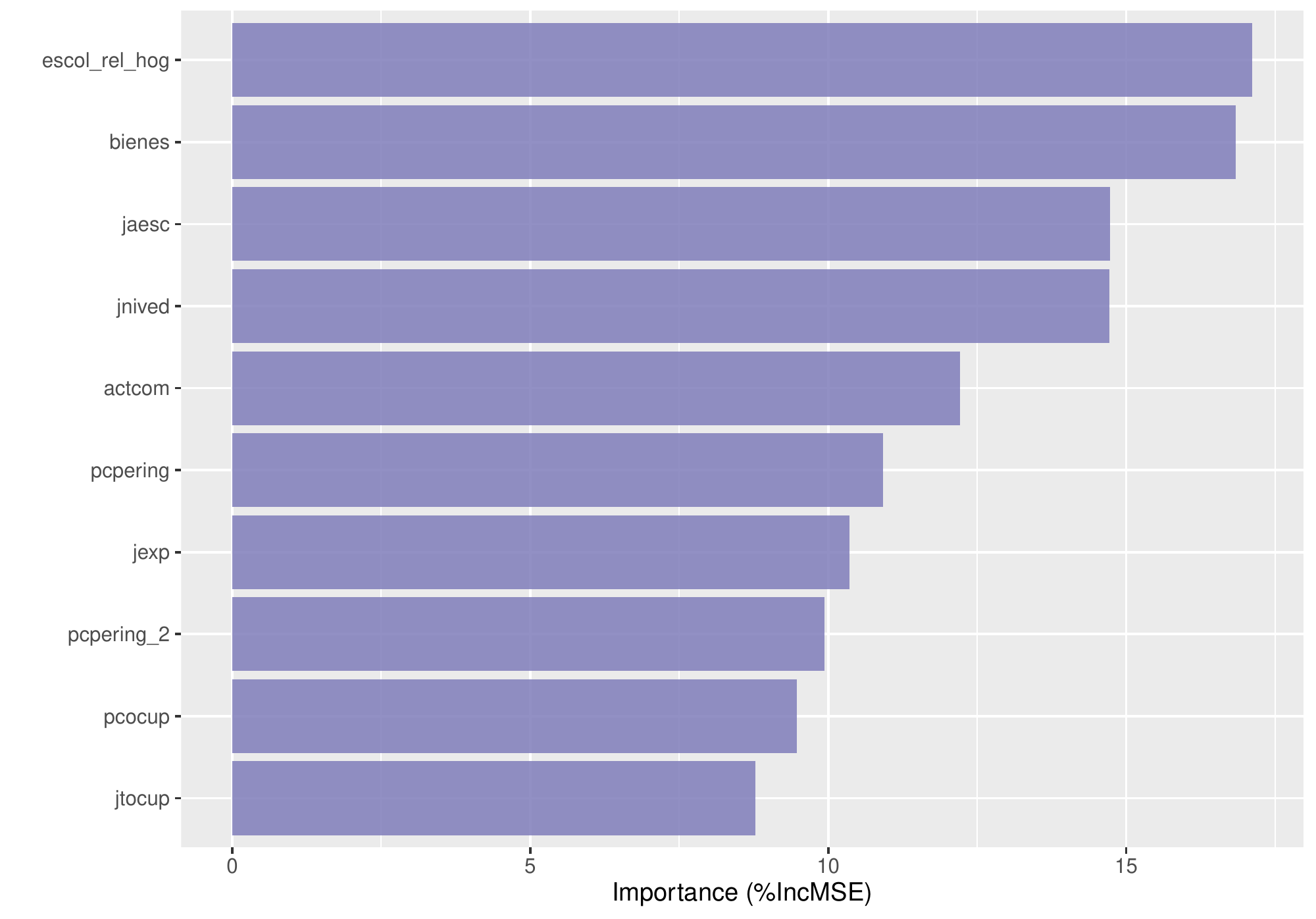}
	\caption{Variable importance in mean decrease accuracy (\%IncMSE) for the ten most influential variables}
	\label{fig:Vipappendix}
\end{figure}

\begin{figure}[ht]
	\centering
	\captionsetup{justification=centering,margin=1.5cm}
	\includegraphics[width=0.95\linewidth]{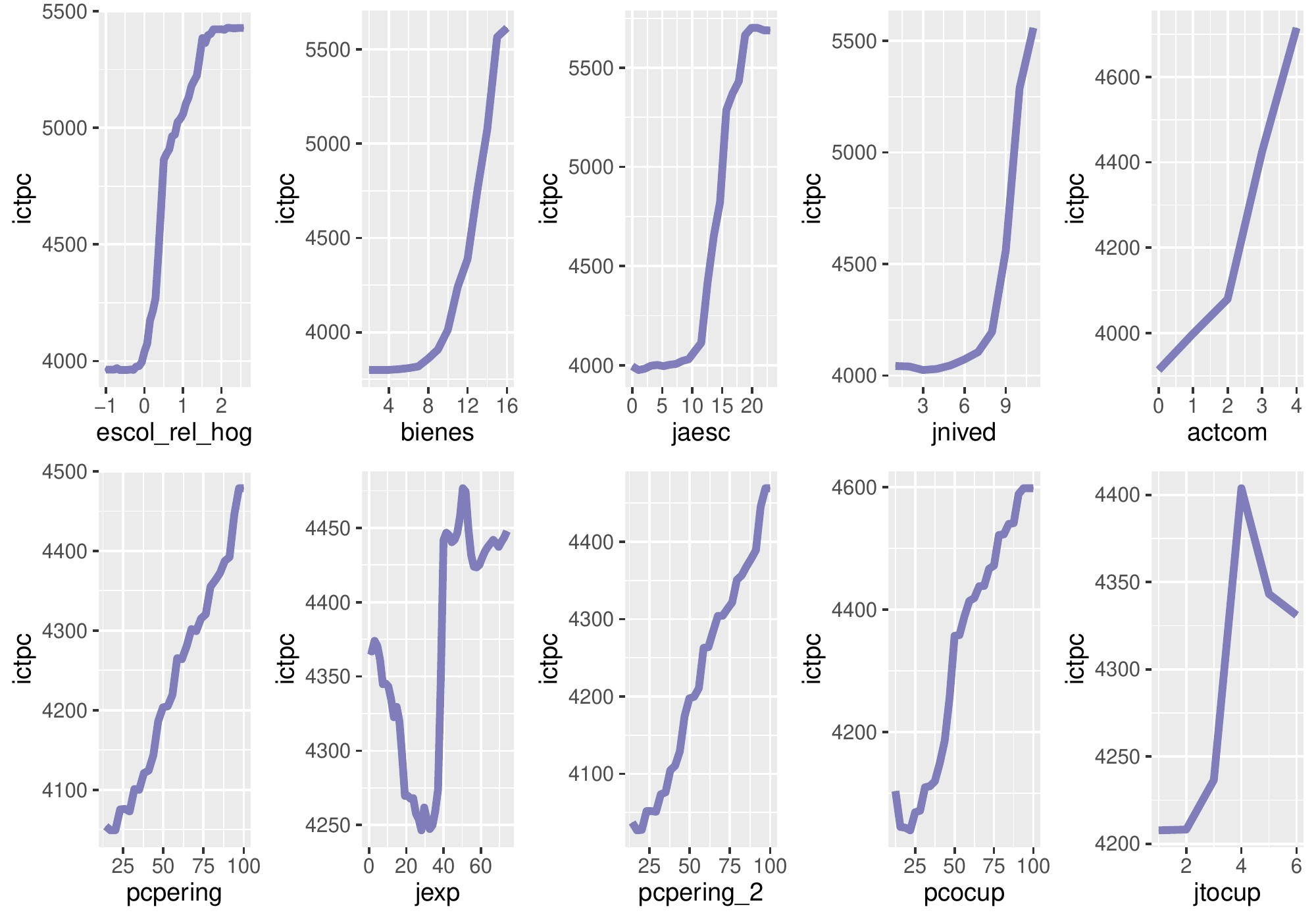}
	\caption{Partial dependence plots for variables ranked by \%IncMSE}
	\label{fig:Pdpappendix}
\end{figure}

The dilemma between predictive precision and the interpretability of random forest models can be mitigated by concepts such as variable importance plots \citep{Greenwell_etal2020} (Figure \ref{fig:Vipappendix}) or an analysis of partial dependence for influential covariates \citep{Greenwell_2017} (Figure \ref{fig:Pdpappendix}). \textcolor{black}{Variable importance is reported by the mean increase in individual mean squared prediction error (\%IncMSE), resulting from the exclusion of the corresponding variable. Partial dependence plots depict the estimated marginal effect of a particular variable on the predicted individual target variables. From the inspection of Figure \ref{fig:Pdpappendix}, we can infer whether relationships between ictpc and predictive variables are monotonic or more complex. Figure \ref{fig:Vipappendix} reveals that the most important variable for the random forest model is the average relative amount of schooling (escol\_rel\_hog), followed by the availability of goods in the household (bienes) as well as the average years of schooling of persons (jaesc). Table \ref{tab:appExplain} in the Appendix B provides explanations on further variables. The most influential variables are related to education, work experience and employment and household assets. Figure \ref{fig:Pdpappendix} indicates rather complex and non-linear relationships between ictpc and its predictive covariates except for two variables related to the number of income earners in the household (pcpering, pcpering\_2).}

We monitor the convergence of the proposed MERF algorithm under a precision of $1e^{-5}$ in relative difference of the GLL criterion and keep the default of $500$ trees. A parameter optimization based on 5-fold cross-validation on the original survey-sample advices the use of $3$ variables at each split for the forest. For the MSE-bootstrap procedure, we use $B=200$.

\begin{figure}[!htb]	
	\centering
	\captionsetup{justification=centering,margin=1.5cm}
	\includegraphics[width=0.95\linewidth]{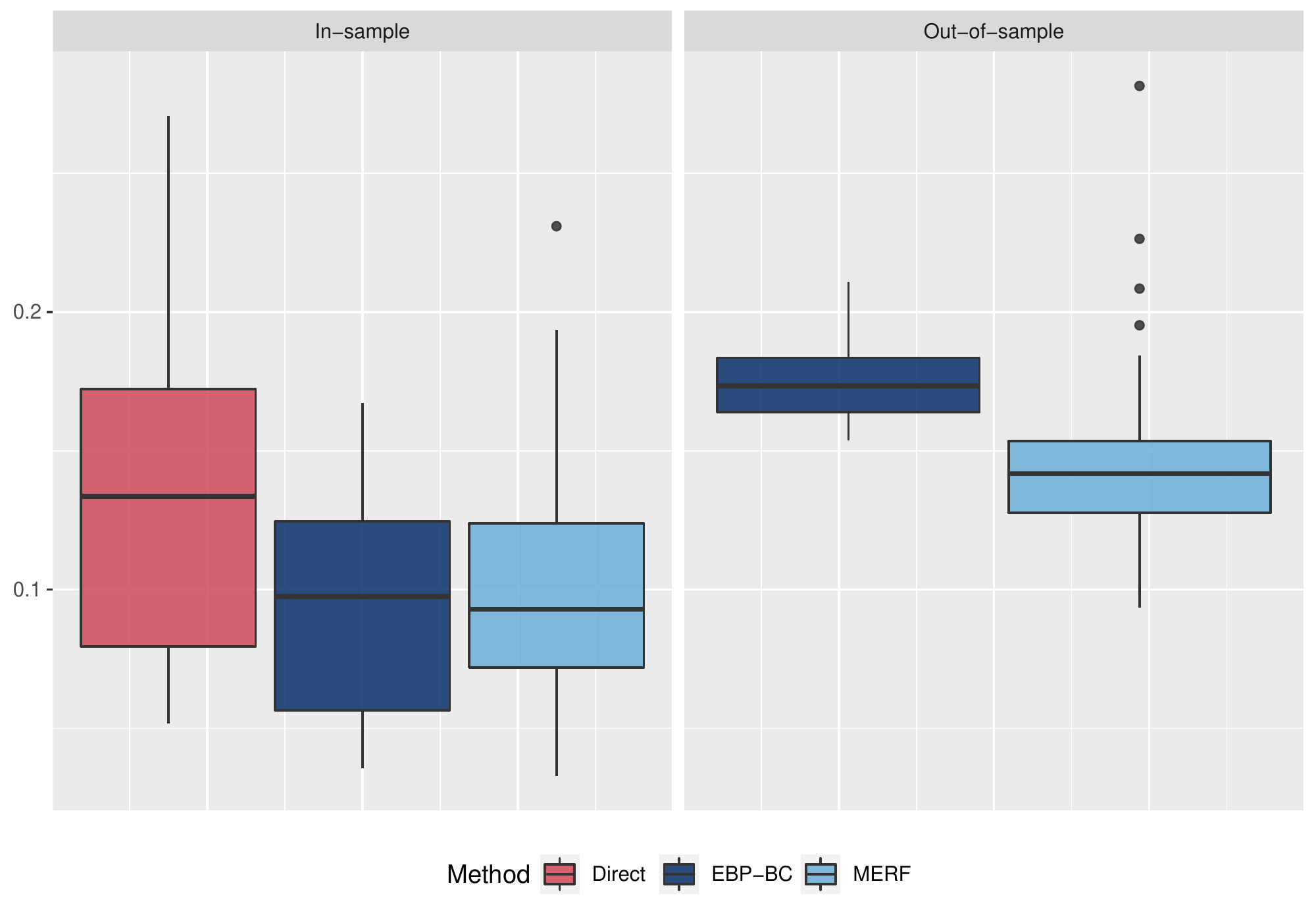}
	\caption{Domain-specific CVs for target variable \textit{ictpc} for in- and out-of-sample domains}
	\label{fig:detailCV}
\end{figure}

Figure \ref{fig:detailCV} reports corresponding CVs for in- and out-of-sample domains. We observe a significant improvement for in-sample CVs of the EBP-BC and the MERF compared to the CVs for direct estimates. CVs of MERFs are slightly lower in median terms than the results for the EBP-BC. However, there exists one outlying area for MERFs. Going into details, the corresponding area of General Zaragoza features no obvious data-specific irregularities, such as extremely low sample size. Nevertheless, General Zaragoza is one of the poorest regions according to our analysis. In the design-based simulation in Section \ref{sec:5.3}, we will pay special attention to differences between MERFs and the EBP-BC regarding their ability to handle comparably extreme estimates given a broad range of relatively high and relative low income areas.

\begin{figure}[ht]
	\centering
	\captionsetup{justification=centering,margin=1.5cm}
	\includegraphics[width=0.95\linewidth]{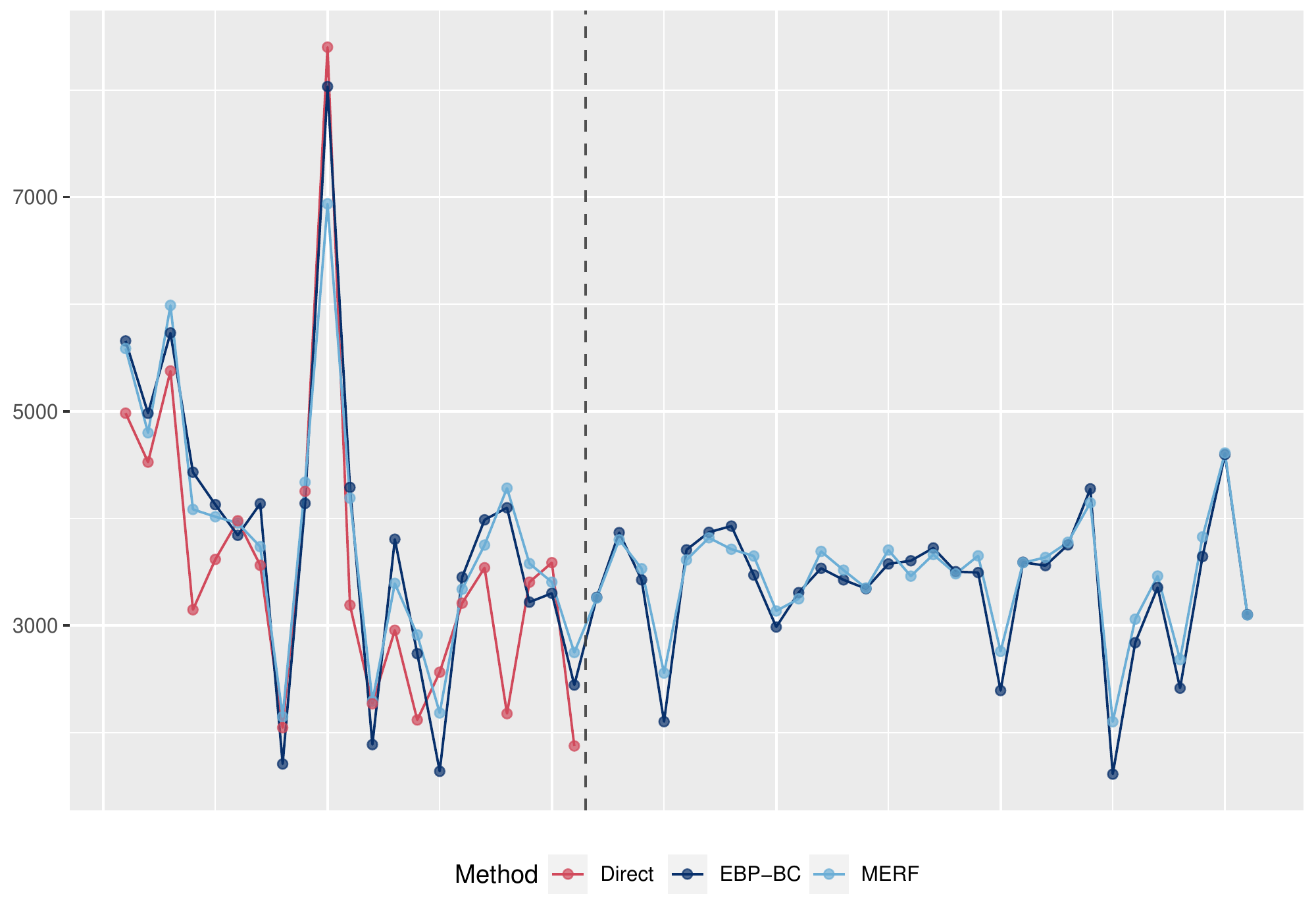}
	\caption{Detailed comparison of point estimates for the domain-level average total household income. The dotted line separates sampled from non-sampled areas. In-sample areas are sorted by decreasing survey-sample size}
	\label{fig:pointEstim}
\end{figure}

Regarding the CVs for out-of-sample areas, we discover an evident advantage of CVs from our proposed MERF approach. From the scrutiny of individual CV values it remains unclear, whether the asset of improved results from MERFs roots in superior point estimates for domain-level averages or its relatively lower MSE-estimates. Figure \ref{fig:pointEstim} compares direct estimates to the model-based estimates for in and out-of-sample domains. Apparently, there exist no systematic differences between the estimates from the EBP-BC and the MERF. Thus, it appears as if the variance of MERF predictions is generally lower. This conjecture is in line with the theoretical properties of random forests \citep{Breiman2001, Biau_Scornet2016}.
In-sample areas in Figure \ref{fig:pointEstim} are sorted by survey-sample sizes. In comparison to the direct estimates, predicted averages of the EBP-BC as well as of the MERF seem less extreme. The obvious irregularity in terms of high-income, is a distinct part of the Monterrey metropolitan area: San Pedro Garza García registers several headquarters of national and international corporations. This economic peculiarity apparently transfers to the income of its citizenship. Figure \ref{fig:mapincome}, underlines the existence of an apparent high-income cluster in this region. Overall, it is interesting to observe how reliable estimates on fine spatial resolution unveil patterns of regional income segregation. Our proposed method of MERFs provides useful results with remarkably higher accuracy than direct estimates and the EBP-BC for most of out-of-sample domains. The following design-based simulation will strengthen the reliability of results and enable an in-depth discussion of our methods for point and MSE-estimates.

\FloatBarrier

\subsection{Evaluation using design-based simulation}\label{sec:5.3}

The design-based simulation allows to directly juxtapose the performance of the proposed MERF-approach to existing SAE-methods for the estimation of area-level means based on empirical data. In this sense, the design-based simulation adds not only insights to the results from the model-based simulation in Section \ref{sec:4}, but also evaluates results from the example in the previous Section \ref{sec:5.3}. We focus on area-level mean-estimates of household income from work \textit{inglabpc} in the Mexican state Nuevo León. As we use the same data with a different target variable, sample and census data properties are similar to the previous example with details provided in Table \ref{tab:Apdetails}. Implementing the design-based simulation, we sample $T=500$ independent samples from the fixed population of our census dataset. Each pseudo-survey-sample mirrors the characteristics of the original survey, as we keep the number of in-sample households similar to the original sample sizes and abstain from sampling out-of-sample municipalities. As a result, we use $500$ equally structured pseudo-survey-samples with equal overall sample size. True values, are defined as the domain-level averages of household incomes from work in the original census.

We consider the same methods as in the model-based simulation in Section \ref{sec:4}. Comparably to Section \ref{sec:5.2}, we use the same working-model for the BHF, the EBP, \textcolor{black}{the EBP-BC and P-SPLINES and} assume it to be fixed throughout the design-based simulation. For the EBP-BC and the MERF, we keep the parameters as already discussed in Section \ref{sec:5.2}.

\begin{figure}[ht]
	\centering
	\captionsetup{justification=centering,margin=1.5cm}
	\includegraphics[width=1\linewidth]{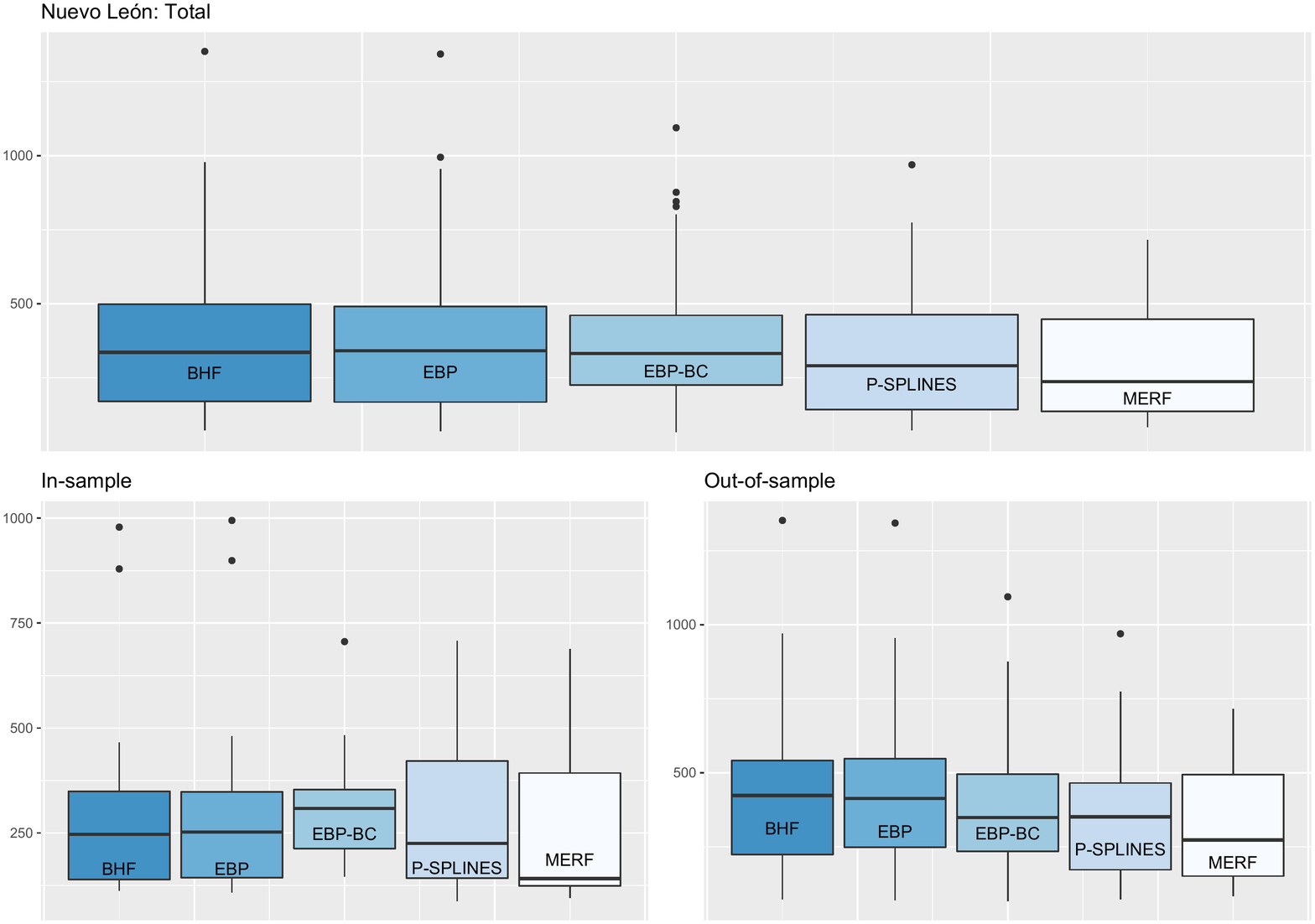}
	\caption{Performance of area-specific point estimates including details on in- and out-of-sample areas. Comparison of empirical RMSEs from the design-based simulation for target variable \textit{inglabpc}}
	\label{fig:DBpoint}
\end{figure}

\begin{table}[!h]
	\centering
	\captionsetup{justification=centering,margin=1.5cm}
	\caption{Mean and Median of RB and RRMSE over in- and out-of-sample areas for point estimates}
	\begin{tabular}{@{\extracolsep{5pt}} lrcccccccc}
		\\[-1.8ex]\hline
		\hline \\[-1.8ex]
		& &\multicolumn{2}{c}{Total} &\multicolumn{2}{c}{In-sample}&\multicolumn{2}{c}{Out-of-sample} \\
		\hline \\[-1.8ex]
		& & Median & Mean & Median & Mean & Median & Mean \\
		\hline \\[-1.8ex]
		\multicolumn{7}{l}{RB[\%]}\\
		\hline \\[-1.8ex]
		&BHF & $14.200$ & $17.800$ & $5.070$ & $11.200$ & $21.800$ & $22.400$ \\
		&EBP & $14.700$ & $17.900$ & $5.310$ & $11.700$ & $22.600$ & $22.300$ \\
		&EBP-BC & $9.650$ & $18.400$ & $6.140$ & $7.150$ & $15.800$ & $26.300$ \\
		&P-SPLINES & $8.120$ & $18.200$ & $0.014$ & $14.200$ & $13.500$ & $21.000$ \\
		&MERF & $7.720$ & $18.600$ & $3.490$ & $17.200$ & $10.600$ & $19.600$ \\
		\hline \\[-1.8ex]
		\multicolumn{7}{l}{RRMSE[\%]}\\
		\hline \\[-1.8ex]
		&BHF & $14.900$ & $21.600$ & $9.520$ & $17.000$ & $23.000$ & $24.800$ \\
		&EBP & $15.900$ & $21.700$ & $9.480$ & $17.400$ & $22.900$ & $24.700$ \\
		&EBP-BC & $14.000$ & $23.900$ & $12.900$ & $15.200$ & $16.100$ & $29.900$ \\
		&P-SPLINES & $11.600$ & $22.800$ & $7.360$ & $20.000$ & $15.000$ & $24.700$ \\
		&MERF & $9.430$ & $21.200$ & $6.130$ & $19.900$ & $12.500$ & $22.100$ \\ 			
		\hline \\[-1.8ex]
	\end{tabular}
\label{tab:DBpoint}
\end{table}

The discussion of results, starts with an investigation into the performance of point estimates. Figure \ref{fig:DBpoint} reports the average RMSE of the area-level mean-estimates for Nuevo León in total and with details on the $21$ in- and $30$ out-of sample areas. The corresponding summary statistics for Figure \ref{fig:DBpoint} are given in Table \ref{tab:MBappendix} in Appendix B. Regarding the total of $51$ areas, we observe no remarkable difference in the performance of the BHF and the EBP, whereas the EBP-BC has lower RMSE on average. \textcolor{black}{P-SPLINES outperform the BHF, EBP and EBP-BC in mean and median terms of RMSE.} The MERF point estimates indicate the lowest RMSEs among all areas resulting in an more than $22$ percent improvement compared to the BHF. Referring to the RMSE for in-sample areas, we see two different ways how the EBP-BC and \textcolor{black}{the adaptive methods of P-SPLINES and MERF} deal with high and unbalanced variation in our true values for certain areas, ranging from $475$ to  about $4004$ pesos: overall, the MERF deals best in modelling the complex survey data and produces highly accurate estimates for the majority of in-sample areas. A closer look at the results, reveals however, that higher RMSE values due to overestimation mainly occur in two areas, both characterised by a very low level of income ($622$ and $544$ pesos respectively). \textcolor{black}{A similar observation can be made for P-SPLINES, although the MERF appears to reproduce the predictive relations more efficiently.} In contrast, we observe the in-sample behaviour of the EBP-BC, which clearly opposes its superior overall performance. The EBP-BC appears to balance extreme estimates by producing slightly worse estimates for each individual in-sample areas, than allowing for individually inferior estimates for specific `outlier'-areas. This behaviour is conceptually rooted in its data-driven transformation-approach. Nevertheless, this property enables the EBP-BC to identify a model, providing stable and precise estimates on the majority of areas, especially the 30 non-sampled areas. Given the data-scenario of Nuevo León, the performance on the out-of-sample areas, delineates each method's quality and stability. In this case, the \textcolor{black}{EBP-BC and the non-parametric approaches of P-SPLINES and MERF} outperform the `traditional' methods (BHF and EBP) in terms of lower RMSE. \textcolor{black}{The median of RMSE of P-SPLINES aligns to the values of the EBP-BC, although the RMSEs of P-SPLINES are lower in means.} One distinct advantage of the MERF is its adaptability and implicit model-selection that is rewarded in the presence of complex data-scenarios.

The findings from Figure \ref{fig:DBpoint}, are strengthened by a discussion of mean and median values of RB and RRMSE in Table \ref{tab:DBpoint}. Referring to all $51$ areas, the RB of the data-driven method of EBP-BC\textcolor{black}{, P-SPLINES} and the MERF is smaller in median values than the RB of BHF and the EBP. Respectively, the MERF shows comparatively low levels of median RB while mean values lie in the same range with competing methods. The obvious difference between mean and median values, indicates the previously discussed existence of inferior estimates for specific regions due to the empirical properties of our underlying data. For the $20$ in-sample areas, \textcolor{black}{P-SPLINES} perform superior to competing methods regarding the median values of RB. The close relation between the mean and median values of RB for the EBP-BC highlight the mentioned balancing-property of the EBP-BC. For the majority of areas in the model-based simulation, i.e. the $30$ non-sampled areas, the EBP-BC\textcolor{black}{, P-SPLINES} as well as the MERF exhibit a comparatively low level of median RB. Especially the MERF captivates by the lowest values in median RRMSE compared to its competitors\textcolor{black}{, while mean values for all cases are within a compatible range}.

Finally, we focus on the performance of the proposed non-parametric MSE-bootstrap procedure. While, the model-based simulation in Section \ref{sec:4} indicates unbiasedness of the proposed bootstrap-scheme under all four scenarios, our results from the design-based simulation require a deeper discussion. \textcolor{black}{We abstain from a discussion of result from our analytical approximation to the area-level MSE because it is limited to in-sample areas and was solely used to contextualize the quality of our proposed bootstrap-scheme from Section \ref{sec:3}.} Table \ref{tab:DBmse} reports the results of RB-RMSE and the RRMSE-RMSE for the corresponding estimates. Figure \ref{fig:DBappendix} in the Appendix B visualizes details from Table \ref{tab:DBmse}. First of all, the values of RRMSE-RMSE are comparable to the most complex scenario in the model-based scenario in Section \ref{sec:4}. The RB-RMSE for the in-sample areas indicates unbiasedness in median terms and an acceptable overestimation regarding the mean RB-RMSE. For out-of-sample areas, we face a moderate underestimation regarding the median value and over-estimation according to mean values. Nevertheless, Figure \ref{fig:DBappendix} in Appendix B reveals, that the mixed signal between mean and median in Table \ref{tab:DBmse} is explained by a balanced mix of under- and over-estimation. Overall, the expectations towards the MSE-bootstrap procedure, given the challenging conditions of this design-based simulation, are met. Especially, the results for in-sample areas, combined with insights from the model-based simulation, indicate a solid and reliable performance of the proposed non-parametric bootstrap procedure. Although, the RB-RMSE for all $51$ areas is driven by the results from out-of-sample areas, the median RB-RMSE is acceptable. Apparently, the MSE-estimates mirror the high variation in sample sizes paired with high and dis-proportional variation of high-income and low-income regions between the $21$ in-sample and $30$ out-of-sample areas. From an applied perspective, the MSE-estimates for out-of-sample areas are nevertheless practicable for the construction of confidence intervals, with a median coverage of $0.97$.

\begin{table}[!h]
	\centering
	\captionsetup{justification=centering,margin=1.5cm}
	\caption{Performance of MSE-estimator in design-based simulation: mean and median of RB and RRMSE over in- and out-of-sample areas}
	\begin{tabular}{@{\extracolsep{5pt}} lcccccccc}
		\\[-1.8ex]\hline
		\hline \\[-1.8ex]
		&\multicolumn{2}{c}{Total} &\multicolumn{2}{c}{In-sample}&\multicolumn{2}{c}{Out-of-sample} \\
		\hline \\[-1.8ex]
		& Median & Mean & Median & Mean & Median & Mean \\
		\hline \\[-1.8ex]
		RB-RMSE[\%] & $$-$1.260$ & $14.300$ & $0.719$ & $7.820$ & $$-$9.460$ & $18.800$ \\
		RRMSE-RMSE[\%] & $48.100$ & $55.900$ & $41.400$ & $47.700$ & $50.900$ & $61.700$ \\
		\hline \\[-1.8ex]
	\end{tabular}
\label{tab:DBmse}
\end{table}

\FloatBarrier

\section{Concluding remarks}\label{sec:6}
In this paper, we explore the potential of tree-based machine learning methods for the estimation of SAE-means. In particular, we provide a solid framework easing the use of random forests for regression within the existing methodological framework of SAE. We highlight the potential of our approach to meet modern requirements of SAE, including the robustness of random forests against model-failure and the applicability for high-dimensional problems processing Big Data sources. The methodological part focusses on the MERF-procedure \citep{Hajjem2014} and implicitly discusses a semi-parametric unit-level mixed model, treating LMM-based SAE-methods, such as the BHF and the EBP, as special cases. The model is fit by an algorithm resembling the EM-algorithm, allowing for flexibility in the specification to model fixed effects as well as random-effects. The proposed point estimator for area-level means is complemented by the non-parametric MSE-bootstrap scheme, building on the REB-bootstrap by \citet{Chambers_Chandra2013} and the bias-corrected estimate for the residual variance by \citet{Mendez_Lohr2011}. We evaluate the performance of point- and MSE-estimates compared to `traditional' SAE-methods by model- and design-based simulations and provide a distinctive SAE example using income data from the Mexican state Nuevo Leòn in Section \ref{sec:5.2}. The model-based simulation in Section \ref{sec:4} demonstrates the ability of point estimates to perform compatibly in classical scenarios and outperform `traditional' methods in the existence non-linear interactions between covariates. The design-based simulation in Section \ref{sec:5.3} confirms the adequacy of MERFs for point estimates under searingly realistic conditions. The model- and design-based simulations indicate that the proposed approach is robust against distributional violations of normality for the random effects and for the unit-level error terms. Concerning our proposed MSE-bootstrap scheme, we conclude its reliability based on the performance in the model-based simulation in Section \ref{sec:4}. Furthermore, we obtain reasonable support for the performance in the application in Section \ref{sec:5.2} and the following design-based simulation in Section \ref{sec:5.3}.

We motivate three major dimensions for further research, including theoretical work, aspects of generalizations and advanced applications using Big Data covariates: from a theoretical perspective, further research is needed to investigate the construction \textcolor{black}{and theoretical discussion of a partial-analytical MSE for area-level means. A conducive strategy is an extension based on our theoretical discussion in the online supplementary materials. Additionally the deduction of recent theoretical results, such as conditions for the consistency of unit-level predictions \citep{Scornet_etal2015} or considerations of individual predictions intervals \citep{wager_etal2014, Zhang2019}, towards area-level indicators, bears potential.} Alternatively, considerations concerning a fully non-parametric formulation of model (\ref{mod1}) impose an interesting research direction. From a survey statistical perspective, our proposed method currently abstains from the use of survey weights which bears a risk if the assumption of non-informative sampling is violated. Nevertheless, there exist approaches incorporating weights into random forests \citep{Winham_etal2013}. The transfer of such ideas to the proposed method of MERFs is subject to ongoing research. Regarding additional generalizations of the proposed method, we aim to extend the use of MERFs towards the estimation of small area quantiles and other non-linear indicators, such as Gini-coefficients or Head Count Ratios. Furthermore, a generalization towards binary or count data is possible and left to further research. The semi-parametric composite formulation of model (\ref{mod1}) allows for $f()$ to adapt any functional form regarding the estimation of the conditional mean of $y_i$ given $X_i$ and technically transfers to other machine learning methods, such as gradient-boosted trees or support vector machines. In terms of advanced applications, we propose the use of MERFs for complex random effect and covariance-structures in empirical problems to the SAE-research community. Equally interesting is the use of high dimensional supplementary data, i.e. Big Data covariates, for the estimation of area-level means, that can be directly handled by the proposed MERF-framework.

\section*{Acknowledgements}
The authors are grateful to CONEVAL for providing the data used in empirical work. The views set out in this paper are those of the authors and do not reflect the official opinion of CONEVAL. The numerical results are not official estimates and are only produced for illustrating the methods.
Additionally, the authors would like to thank the HPC Service of ZEDAT, Freie Universität Berlin, for computing time.

\section*{Appendix A}

After convergence of the algorithm introduced in Section \ref{sec:2.2}, we obtain an optimal non-parametric estimator for $\hat{f}()$. In the following, we facilitate the notation and refer to $\hat{f}^{OOB}()$ simply as $\hat{f}()$. The best predictor for the random effects for known parameters $H_i$ and $R_i$ must maximize the generalized log-likelihood criterion:
$$GLL (f,v_i | y) = \sum_{i=1}^{D}\{ [y_i - f(X_i) - Z_i v_i ]' R_i^{-1} [ y_i - f(X_i) - Z_i v_i] + v_i ' H^{-1} v_i +log |H|+ log|R_i|\}.$$
Finding a maximum for $GLL$ is equivalent to the problem of finding a maximizer for $v$ in the first term of the summation of the proposed GLL-criterion:
$$[y - \hat{f}(X_i) - Z_i v_i ]' R_i^{-1} [ y_i - \hat{f}(X_i) - Z_i v_i] + v_i ' H_i^{-1} v_i.$$

Reshaping leads to:
\begin{flalign*}
	[y_i - \hat{f}(X_i) - Z_i v_i ]' R_i^{-1}[ y_i - \hat{f}(X_i) - Z_i v_i] + v_i ' H_i^{-1} v_i &= \\
	y_i' R_i^{-1}y_i - y_i' R_i^{-1}\hat{f}(X_i) -y_i' R_i^{-1} Z_i v_i -\hat{f}(X_i)'R_i^{-1}y_i + \hat{f}(X_i)'R_i^{-1}\hat{f}(X_i) &+ \\ \hat{f}(X_i)'R_i^{-1}Z_i v_i -  (Z_i v_i)'R_i^{-1}y_i+(Z_i v_i)'R_i^{-1}\hat{f}(X_i)+(Z_i v_i)'R_i^{-1}(Z_i v_i)+v_i ' H_i^{-1} v_i
\end{flalign*}

Now we derive the expression with respect to $v$ and set it to 0 in order to find the maximizer

\begin{flalign*}
	-y_i' R_i^{-1} Z_i+ \hat{f}(X_i)'R_i^{-1} Z_i -Z_i'R_i^{-1}y_i+\\ Z_i'R_i^{-1}\hat{f}(X_i)+Z_i'R_i^{-1}Z_i v_i+(Z_i v_i)'R_i^{-1}y_i+Z_i+ 2 H_i^{-1} v_i &= 0 &\Longleftrightarrow\\
	-2 y_i R_i^{-1} Z_i + 2 Z_i'R_i^{-1}\hat{f}(X_i) + 2 Z_i'R_i^{-1}Z_i v_i +2 H_i^{-1} v_i  &= 0 &\Longleftrightarrow \\
	- y_i R_i^{-1} Z_i + Z_i'R_i^{-1}\hat{f}(X_i) + Z_i'R_i^{-1}Z_i v_i + H_i^{-1} v_i  &= 0 &\Longleftrightarrow\\
	y_i R_i^{-1} Z_i - Z_i'R_i^{-1}\hat{f}(X_i) &= Z_i'R_i^{-1}Z_i v_i + H_i^{-1} v_i &\Longleftrightarrow\\
	y_i R_i^{-1} Z_i - Z_i'R_i^{-1}\hat{f}(X_i) &= (Z_i'R_i^{-1}Z_i + H_i^{-1}) v_i &\Longleftrightarrow\\
	(Z_i'R_i^{-1}Z_i + H_i^{-1})^{-1} (Z_i'R_i^{-1} (y_i - \hat{f}(X_i)) &= v_i
\end{flalign*}
\begin{flalign*}
	v_i &= (Z_i'R_i^{-1}Z_i + H_i^{-1})^{-1} (Z_i'R_i^{-1} (y_i - \hat{f}(X_i))\\
	&= (Z_i'R_i^{-1}Z_i + H_i^{-1})^{-1} V_i V_i^{-1} (Z_i'R_i^{-1} (y_i - \hat{f}(X_i)\\
	&= (Z_i'R_i^{-1}Z_i + H_i^{-1})^{-1} Z_i'R_i^{-1} (R_i+Z_i H_i Z_i')V_i^{-1}(y_i - \hat{f}(X_i))\\
	&= (Z_i'R_i^{-1}Z_i + H_i^{-1})^{-1} (Z_i' +Z_i'R_i^{-1}Z_i H_i Z_i')V_i^{-1}(y_i - \hat{f}(X_i))\\
	&= (Z_i'R_i^{-1}Z_i + H_i^{-1})^{-1}(Z_i'R_i^{-1}Z_i + H_i^{-1}) (H_i Z_i'V_i^{-1}(y_i - \hat{f}(X_i)))\\
&=(H_i Z_i'V_i^{-1}(y_i - \hat{f}(X_i)))
\end{flalign*}

The solution of the maximization problem is given by $\hat{v_i}^{*} = H_i Z_i'V_i^{-1}(y_i - \hat{f}(X_i))$. Note, for $\hat{f}(X_i) = X_i\hat{\beta}$, the optimality solution resembles the BLUP.

\section*{Appendix B: additional simulation results and model-diagnostics}

\begin{figure}[ht]
	\centering
	\captionsetup{justification=centering,margin=1.5cm}
	\includegraphics[width=1\linewidth]{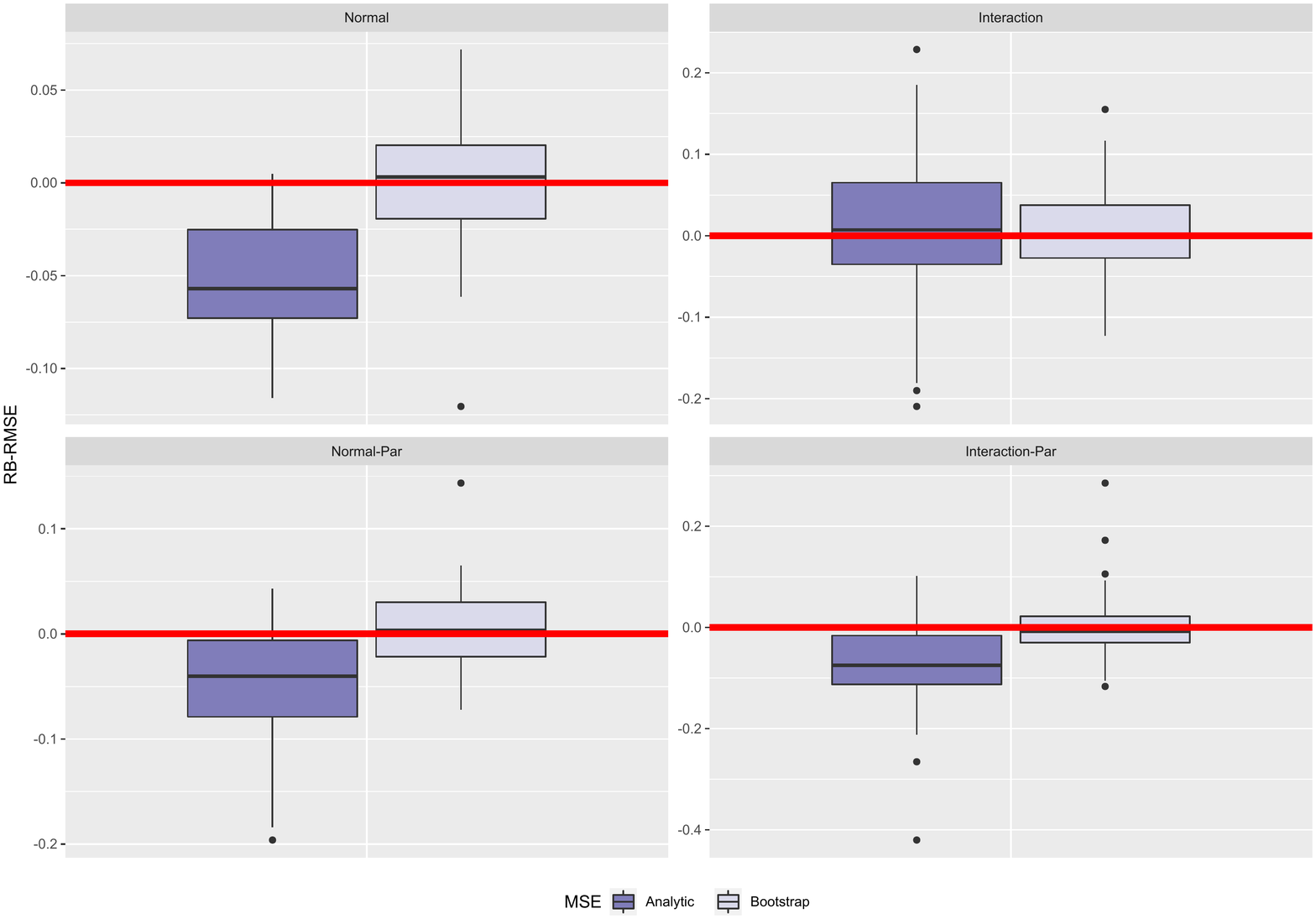}
	\caption{Details on the performance of the proposed bootstrap MSE-estimator and the analytic approximation in the model-based simulation: boxplots of the area-specific RB-RMSEs averaged over simulation runs}
	\label{fig:MBappendix}
\end{figure}

\begin{table}[ht]
	\centering
	\captionsetup{justification=centering,margin=1.5cm}
	\caption{Explanation of most influential variables according to the random forest model in the application of Section \ref{sec:5.2}}
\begin{tabular}{@{\extracolsep{1pt}} lr}
	\\[-1.8ex]\hline
	\hline \\[-1.8ex]
	Variable name & Explanation \\
	\hline
	ictpc & Total household income per capita \\
	escol\_rel\_hog & Average relative amount of schooling standardized \\
	& by age and sex of household members \\
	bienes & Availability of goods in the household \\
	jaesc & Average years of schooling of persons in the household \\
	jnived & Formal education of the head of the household \\
	actcom & Assets in the household \\
	pcpering & Percentage of income earners in the household \\
	jexp & Years of working experience of the head of the household \\
	pcpering\_2 & Number of income earners in the household by household size \\
	pcocup & Percentage of people employed in the household \\
	jtocup & Occupation type \\
	\hline
\end{tabular}
\label{tab:appExplain}
\end{table}

\begin{table}[ht]
	\centering
	\captionsetup{justification=centering,margin=1.5cm}
	\caption{Performance of point estimates in design-based simulation: summary statistics of empirical RMSE for area-level mean-estimates}
	\begin{tabular}{@{\extracolsep{5pt}} lrcccccccc}
		\\[-1.8ex]\hline
		\hline \\[-1.8ex]
		Areas & Method & Min & 1st Qu. & Median & Mean & 3rd Qu. & Max \\
		\hline
		\multirow{ 4}{*}{Total} & BHF & 72.58 & 170.63 & 336.06 & 386.40 & 498.13 & 1351.82 \\
		 & EBP & 68.83 & 168.13 & 341.10 & 387.91 & 490.14 & 1342.92 \\
		& EBP-BC & 65.22 & 225.84 & 331.64 & 376.49 & 460.86 & 1094.08 \\
		& P-SPLINES & 72.70 & 142.91 & 290.84 & 337.11 & 462.94 & 969.15 \\
		 & MERF & 82.58 & 136.01 & 236.86 & 298.20 & 447.21 & 716.72 \\
		 \hline
		\multirow{ 4}{*}{In-sample} & BHF & 111.93 & 139.13 & 246.62 & 301.90 & 349.26 & 978.49 \\
		& EBP & 107.41 & 143.07 & 251.95 & 308.47 & 348.00 & 994.45 \\
		& EBP-BC & 145.48 & 212.64 & 308.43 & 314.08 & 353.02 & 705.81 \\
		& P-SPLINES & 86.69 & 142.70 & 224.82 & 285.37 & 421.29 & 707.71 \\
	    & MERF & 94.56 & 123.72 & 141.16 & 264.27 & 392.73 & 688.74 \\
	    \hline
		\multirow{ 4}{*}{Out-of-sample} & BHF & 72.58 & 224.35 & 422.90 & 445.55 & 541.31 & 1351.82 \\
		& EBP & 68.83 & 248.89 & 412.91 & 443.52 & 547.12 & 1342.92 \\
		& EBP-BC & 65.22 & 234.98 & 348.83 & 420.17 & 494.83 & 1094.08 \\
		& P-SPLINES & 72.70 & 172.29 & 351.17 & 373.34 & 465.36 & 969.15 \\
		& MERF & 82.58 & 151.24 & 272.97 & 321.95 & 493.85 & 716.72 \\
		\hline
	\end{tabular}
\label{tab:MBappendix}
\end{table}

\begin{figure}[ht]
	\centering
	\captionsetup{justification=centering,margin=1.5cm}
	\includegraphics[width=1\linewidth]{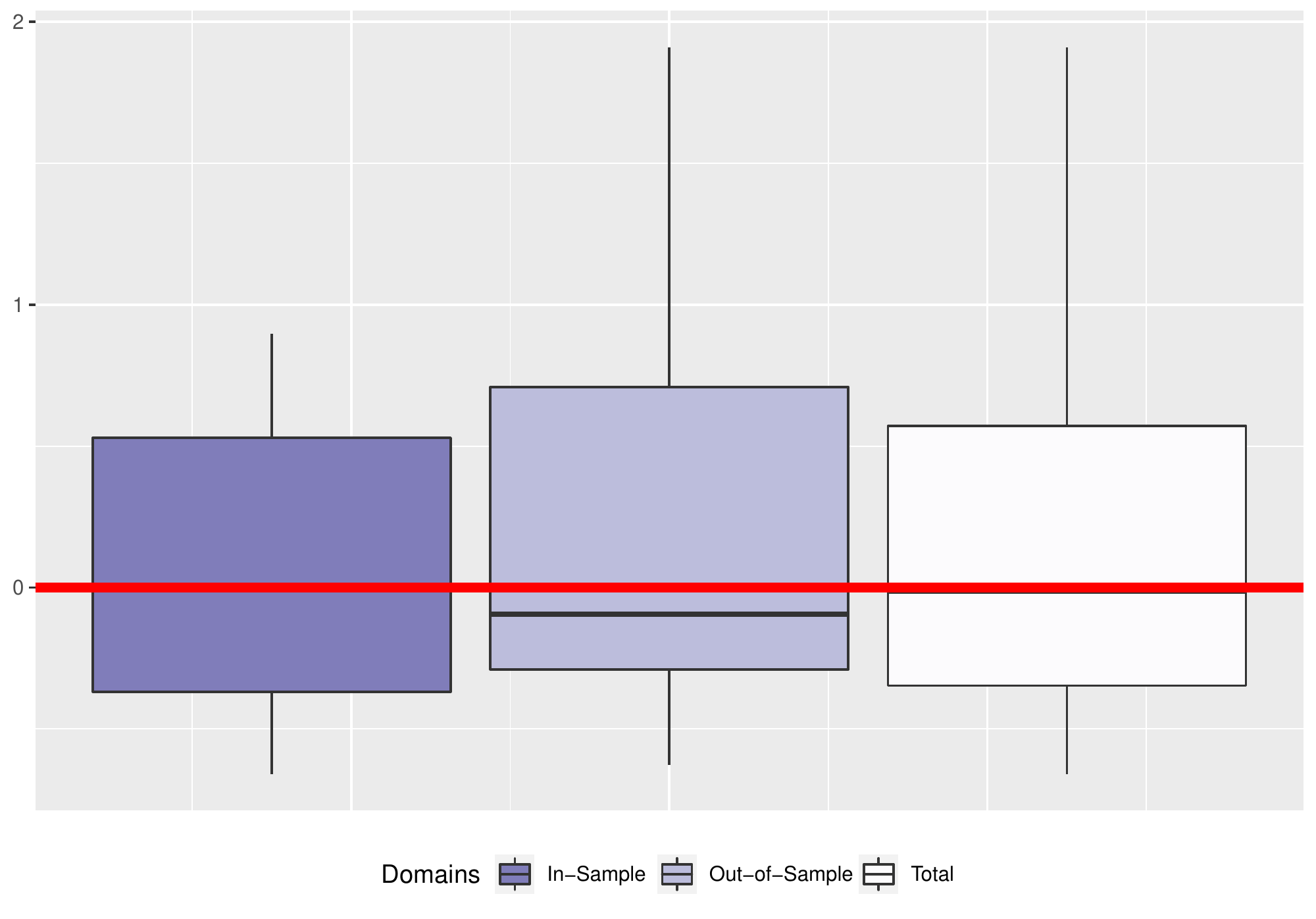}
	\caption{Details on the performance of the proposed MSE-estimator in the design-based simulation: boxplots of the area-specific RB-RMSEs averaged over simulation runs including details on in- and out-of-sample areas}
	\label{fig:DBappendix}
\end{figure}
\clearpage

\bibliographystyle{apacite}				
\bibliography{./MERF_paper_refs_1}

\begin{thebibliography}{}

\bibitem [\protect \citeauthoryear {%
Alfons%
\ \BBA {} Templ%
}{%
Alfons%
\ \BBA {} Templ%
}{%
{\protect \APACyear {2013}}%
}]{%
Alfons_Templ2013}
\APACinsertmetastar {%
Alfons_Templ2013}%
\begin{APACrefauthors}%
Alfons, A.%
\BCBT {}\ \BBA {} Templ, M.%
\end{APACrefauthors}%
\unskip\
\newblock
\APACrefYearMonthDay{2013}{}{}.
\newblock
{\BBOQ}\APACrefatitle {{Estimation of Social Exclusion Indicators from Complex
  Surveys: The R Package laeken}} {{Estimation of Social Exclusion Indicators
  from Complex Surveys: The R Package laeken}}.{\BBCQ}
\newblock
\APACjournalVolNumPages{Journal of Statistical Software}{54}{15}{1--25}.
\PrintBackRefs{\CurrentBib}

\bibitem [\protect \citeauthoryear {%
Anderson%
, Guikema%
, Zaitchik%
\BCBL {}\ \BBA {} Pan%
}{%
Anderson%
\ \protect \BOthers {.}}{%
{\protect \APACyear {2014}}%
}]{%
Anderson_etal2014}
\APACinsertmetastar {%
Anderson_etal2014}%
\begin{APACrefauthors}%
Anderson, W.%
, Guikema, S.%
, Zaitchik, B.%
\BCBL {}\ \BBA {} Pan, W.%
\end{APACrefauthors}%
\unskip\
\newblock
\APACrefYearMonthDay{2014}{}{}.
\newblock
{\BBOQ}\APACrefatitle {{Methods for Estimating Population Density in
  Data-Limited Areas: Evaluating Regression and Tree-Based Models in Peru}}
  {{Methods for Estimating Population Density in Data-Limited Areas: Evaluating
  Regression and Tree-Based Models in Peru}}.{\BBCQ}
\newblock
\APACjournalVolNumPages{PLoS One}{9}{7}{}.
\PrintBackRefs{\CurrentBib}

\bibitem [\protect \citeauthoryear {%
Athey%
, Tibshirani%
\BCBL {}\ \BBA {} Wager.%
}{%
Athey%
\ \protect \BOthers {.}}{%
{\protect \APACyear {2019}}%
}]{%
Athey_etal2019}
\APACinsertmetastar {%
Athey_etal2019}%
\begin{APACrefauthors}%
Athey, S.%
, Tibshirani, J.%
\BCBL {}\ \BBA {} Wager., S.%
\end{APACrefauthors}%
\unskip\
\newblock
\APACrefYearMonthDay{2019}{}{}.
\newblock
{\BBOQ}\APACrefatitle {{Generalized Random Forests}} {{Generalized Random
  Forests}}.{\BBCQ}
\newblock
\APACjournalVolNumPages{The Annals of Statistics}{47}{2}{1148--1178}.
\PrintBackRefs{\CurrentBib}

\bibitem [\protect \citeauthoryear {%
Bates%
, M{\"a}chler%
, Bolker%
\BCBL {}\ \BBA {} Walker%
}{%
Bates%
\ \protect \BOthers {.}}{%
{\protect \APACyear {2015}}%
}]{%
Bates_etal2015}
\APACinsertmetastar {%
Bates_etal2015}%
\begin{APACrefauthors}%
Bates, D.%
, M{\"a}chler, M.%
, Bolker, B.%
\BCBL {}\ \BBA {} Walker, S.%
\end{APACrefauthors}%
\unskip\
\newblock
\APACrefYearMonthDay{2015}{}{}.
\newblock
{\BBOQ}\APACrefatitle {{Fitting Linear Mixed-Effects Models Using lme4}}
  {{Fitting Linear Mixed-Effects Models Using lme4}}.{\BBCQ}
\newblock
\APACjournalVolNumPages{Journal of Statistical Software}{67}{1}{1--48}.
\PrintBackRefs{\CurrentBib}

\bibitem [\protect \citeauthoryear {%
Battese%
, Harter%
\BCBL {}\ \BBA {} Fuller%
}{%
Battese%
\ \protect \BOthers {.}}{%
{\protect \APACyear {1988}}%
}]{%
Battese_etal1988}
\APACinsertmetastar {%
Battese_etal1988}%
\begin{APACrefauthors}%
Battese, G\BPBI E.%
, Harter, R\BPBI M.%
\BCBL {}\ \BBA {} Fuller, W\BPBI A.%
\end{APACrefauthors}%
\unskip\
\newblock
\APACrefYearMonthDay{1988}{}{}.
\newblock
{\BBOQ}\APACrefatitle {{An Error-Components Model for Prediction of County Crop
  Areas Using Survey and Satellite Data}} {{An Error-Components Model for
  Prediction of County Crop Areas Using Survey and Satellite Data}}.{\BBCQ}
\newblock
\APACjournalVolNumPages{Journal of the American Statistical
  Association}{83}{401}{28--36}.
\PrintBackRefs{\CurrentBib}

\bibitem [\protect \citeauthoryear {%
Biau%
\ \BBA {} Scornet%
}{%
Biau%
\ \BBA {} Scornet%
}{%
{\protect \APACyear {2016}}%
}]{%
Biau_Scornet2016}
\APACinsertmetastar {%
Biau_Scornet2016}%
\begin{APACrefauthors}%
Biau, G.%
\BCBT {}\ \BBA {} Scornet, E.%
\end{APACrefauthors}%
\unskip\
\newblock
\APACrefYearMonthDay{2016}{}{}.
\newblock
{\BBOQ}\APACrefatitle {{A Random Forest Guided Tour}} {{A Random Forest Guided
  Tour}}.{\BBCQ}
\newblock
\APACjournalVolNumPages{Test}{25}{2}{197--227}.
\PrintBackRefs{\CurrentBib}

\bibitem [\protect \citeauthoryear {%
Bilton%
, Jones%
, Ganesh%
\BCBL {}\ \BBA {} Haslett%
}{%
Bilton%
\ \protect \BOthers {.}}{%
{\protect \APACyear {2017}}%
}]{%
Bilton_etal2017}
\APACinsertmetastar {%
Bilton_etal2017}%
\begin{APACrefauthors}%
Bilton, P.%
, Jones, G.%
, Ganesh, S.%
\BCBL {}\ \BBA {} Haslett, S.%
\end{APACrefauthors}%
\unskip\
\newblock
\APACrefYearMonthDay{2017}{}{}.
\newblock
{\BBOQ}\APACrefatitle {{Classification Trees for Poverty Mapping}}
  {{Classification Trees for Poverty Mapping}}.{\BBCQ}
\newblock
\APACjournalVolNumPages{Computational Statistics \& Data
  Analysis}{115}{}{53--66}.
\PrintBackRefs{\CurrentBib}

\bibitem [\protect \citeauthoryear {%
Breiman%
}{%
Breiman%
}{%
{\protect \APACyear {1996}}%
}]{%
Breiman1996bagging}
\APACinsertmetastar {%
Breiman1996bagging}%
\begin{APACrefauthors}%
Breiman, L.%
\end{APACrefauthors}%
\unskip\
\newblock
\APACrefYearMonthDay{1996}{}{}.
\newblock
{\BBOQ}\APACrefatitle {{Bagging Predictors}} {{Bagging Predictors}}.{\BBCQ}
\newblock
\APACjournalVolNumPages{Machine Learning}{24}{2}{123--140}.
\PrintBackRefs{\CurrentBib}

\bibitem [\protect \citeauthoryear {%
Breiman%
}{%
Breiman%
}{%
{\protect \APACyear {2001}}%
}]{%
Breiman2001}
\APACinsertmetastar {%
Breiman2001}%
\begin{APACrefauthors}%
Breiman, L.%
\end{APACrefauthors}%
\unskip\
\newblock
\APACrefYearMonthDay{2001}{}{}.
\newblock
{\BBOQ}\APACrefatitle {{Random Forests}} {{Random Forests}}.{\BBCQ}
\newblock
\APACjournalVolNumPages{Machine Learning}{45}{1}{5--32}.
\PrintBackRefs{\CurrentBib}

\bibitem [\protect \citeauthoryear {%
Breiman%
, Friedman%
, Stone%
\BCBL {}\ \BBA {} Olshen%
}{%
Breiman%
\ \protect \BOthers {.}}{%
{\protect \APACyear {1984}}%
}]{%
Breiman_etal1984}
\APACinsertmetastar {%
Breiman_etal1984}%
\begin{APACrefauthors}%
Breiman, L.%
, Friedman, J.%
, Stone, C\BPBI J.%
\BCBL {}\ \BBA {} Olshen, R\BPBI A.%
\end{APACrefauthors}%
\unskip\
\newblock
\APACrefYear{1984}.
\newblock
\APACrefbtitle {{Classification and Regression Trees}} {{Classification and
  Regression Trees}}.
\newblock
\APACaddressPublisher{}{CRC Press}.
\PrintBackRefs{\CurrentBib}

\bibitem [\protect \citeauthoryear {%
Capitaine%
, Genuer%
\BCBL {}\ \BBA {} Thi{\'e}baut%
}{%
Capitaine%
\ \protect \BOthers {.}}{%
{\protect \APACyear {2021}}%
}]{%
capitaine_etal2021}
\APACinsertmetastar {%
capitaine_etal2021}%
\begin{APACrefauthors}%
Capitaine, L.%
, Genuer, R.%
\BCBL {}\ \BBA {} Thi{\'e}baut, R.%
\end{APACrefauthors}%
\unskip\
\newblock
\APACrefYearMonthDay{2021}{}{}.
\newblock
{\BBOQ}\APACrefatitle {{Random Forests for High-Dimensional Longitudinal Data}}
  {{Random Forests for High-Dimensional Longitudinal Data}}.{\BBCQ}
\newblock
\APACjournalVolNumPages{Statistical Methods in Medical
  Research}{30}{1}{166--184}.
\PrintBackRefs{\CurrentBib}

\bibitem [\protect \citeauthoryear {%
Chambers%
\ \BBA {} Chandra%
}{%
Chambers%
\ \BBA {} Chandra%
}{%
{\protect \APACyear {2013}}%
}]{%
Chambers_Chandra2013}
\APACinsertmetastar {%
Chambers_Chandra2013}%
\begin{APACrefauthors}%
Chambers, R.%
\BCBT {}\ \BBA {} Chandra, H.%
\end{APACrefauthors}%
\unskip\
\newblock
\APACrefYearMonthDay{2013}{}{}.
\newblock
{\BBOQ}\APACrefatitle {{A Random Effect Block Bootstrap for Clustered Data}}
  {{A Random Effect Block Bootstrap for Clustered Data}}.{\BBCQ}
\newblock
\APACjournalVolNumPages{Journal of Computational and Graphical
  Statistics}{22}{2}{452--470}.
\PrintBackRefs{\CurrentBib}

\bibitem [\protect \citeauthoryear {%
Chambers%
\ \BBA {} Tzavidis%
}{%
Chambers%
\ \BBA {} Tzavidis%
}{%
{\protect \APACyear {2006}}%
}]{%
Cha06}
\APACinsertmetastar {%
Cha06}%
\begin{APACrefauthors}%
Chambers, R.%
\BCBT {}\ \BBA {} Tzavidis, N.%
\end{APACrefauthors}%
\unskip\
\newblock
\APACrefYearMonthDay{2006}{}{}.
\newblock
{\BBOQ}\APACrefatitle {M-Quantile Models for Small Area Estimation} {M-quantile
  models for small area estimation}.{\BBCQ}
\newblock
\APACjournalVolNumPages{Biometrika}{93}{}{255-268}.
\PrintBackRefs{\CurrentBib}

\bibitem [\protect \citeauthoryear {%
Dagdoug%
, Goga%
\BCBL {}\ \BBA {} Haziza%
}{%
Dagdoug%
\ \protect \BOthers {.}}{%
{\protect \APACyear {2021}}%
}]{%
Dagdougetal2020}
\APACinsertmetastar {%
Dagdougetal2020}%
\begin{APACrefauthors}%
Dagdoug, M.%
, Goga, C.%
\BCBL {}\ \BBA {} Haziza, D.%
\end{APACrefauthors}%
\unskip\
\newblock
\APACrefYearMonthDay{2021}{}{}.
\newblock
{\BBOQ}\APACrefatitle {Model-assisted estimation through random forests in
  finite population sampling} {Model-assisted estimation through random forests
  in finite population sampling}.{\BBCQ}
\newblock
\APACjournalVolNumPages{Journal of the American Statistical
  Association}{}{}{1--18}.
\PrintBackRefs{\CurrentBib}

\bibitem [\protect \citeauthoryear {%
Datta%
\ \BBA {} Lahiri%
}{%
Datta%
\ \BBA {} Lahiri%
}{%
{\protect \APACyear {2000}}%
}]{%
Datta_Lahiri2000}
\APACinsertmetastar {%
Datta_Lahiri2000}%
\begin{APACrefauthors}%
Datta, G\BPBI S.%
\BCBT {}\ \BBA {} Lahiri, P.%
\end{APACrefauthors}%
\unskip\
\newblock
\APACrefYearMonthDay{2000}{}{}.
\newblock
{\BBOQ}\APACrefatitle {{A Unified Measure of Uncertainty of Estimated Best
  Linear Unbiased Predictors in Small Area Estimation Problems}} {{A Unified
  Measure of Uncertainty of Estimated Best Linear Unbiased Predictors in Small
  Area Estimation Problems}}.{\BBCQ}
\newblock
\APACjournalVolNumPages{Statistica Sinica}{10}{2}{613--627}.
\PrintBackRefs{\CurrentBib}

\bibitem [\protect \citeauthoryear {%
De~Moliner%
\ \BBA {} Goga%
}{%
De~Moliner%
\ \BBA {} Goga%
}{%
{\protect \APACyear {2018}}%
}]{%
DeMolinerGoga2018}
\APACinsertmetastar {%
DeMolinerGoga2018}%
\begin{APACrefauthors}%
De~Moliner, A.%
\BCBT {}\ \BBA {} Goga, C.%
\end{APACrefauthors}%
\unskip\
\newblock
\APACrefYearMonthDay{2018}{}{}.
\newblock
{\BBOQ}\APACrefatitle {{Sample-Based setimation of Mean Electricity Consumption
  Curves for Small Domains}} {{Sample-Based setimation of Mean Electricity
  Consumption Curves for Small Domains}}.{\BBCQ}
\newblock
\APACjournalVolNumPages{Survey Methodology}{44}{2}{}.
\PrintBackRefs{\CurrentBib}

\bibitem [\protect \citeauthoryear {%
Diallo%
\ \BBA {} Rao%
}{%
Diallo%
\ \BBA {} Rao%
}{%
{\protect \APACyear {2018}}%
}]{%
DialloRao2018}
\APACinsertmetastar {%
DialloRao2018}%
\begin{APACrefauthors}%
Diallo, M\BPBI S.%
\BCBT {}\ \BBA {} Rao, J\BPBI N\BPBI K.%
\end{APACrefauthors}%
\unskip\
\newblock
\APACrefYearMonthDay{2018}{}{}.
\newblock
{\BBOQ}\APACrefatitle {{Small Area Estimation of Complex Parameters Under
  Unit-Level Models with Skew-Normal Errors}} {{Small Area Estimation of
  Complex Parameters Under Unit-Level Models with Skew-Normal Errors}}.{\BBCQ}
\newblock
\APACjournalVolNumPages{Scandinavian Journal of Statistics}{45}{4}{1092--1116}.
\PrintBackRefs{\CurrentBib}

\bibitem [\protect \citeauthoryear {%
Efron%
\ \BBA {} Hastie%
}{%
Efron%
\ \BBA {} Hastie%
}{%
{\protect \APACyear {2016}}%
}]{%
Efron_Hastie2016}
\APACinsertmetastar {%
Efron_Hastie2016}%
\begin{APACrefauthors}%
Efron, B.%
\BCBT {}\ \BBA {} Hastie, T.%
\end{APACrefauthors}%
\unskip\
\newblock
\APACrefYear{2016}.
\newblock
\APACrefbtitle {{Computer Age Statistical Inference}} {{Computer Age
  Statistical Inference}}.
\newblock
\APACaddressPublisher{}{Cambridge University Press}.
\PrintBackRefs{\CurrentBib}

\bibitem [\protect \citeauthoryear {%
Fay%
\ \BBA {} Herriot%
}{%
Fay%
\ \BBA {} Herriot%
}{%
{\protect \APACyear {1979}}%
}]{%
Fay_Heriot1979}
\APACinsertmetastar {%
Fay_Heriot1979}%
\begin{APACrefauthors}%
Fay, R\BPBI E.%
\BCBT {}\ \BBA {} Herriot, R\BPBI A.%
\end{APACrefauthors}%
\unskip\
\newblock
\APACrefYearMonthDay{1979}{}{}.
\newblock
{\BBOQ}\APACrefatitle {{Estimates of Income for Small Places: An Application of
  James-Stein Procedures to Census Data}} {{Estimates of Income for Small
  Places: An Application of James-Stein Procedures to Census Data}}.{\BBCQ}
\newblock
\APACjournalVolNumPages{Journal of the American Statistical
  Association}{74}{366a}{269--277}.
\PrintBackRefs{\CurrentBib}

\bibitem [\protect \citeauthoryear {%
Gonz{\'a}lez-Manteiga%
, Lombard{\'\i}a%
, Molina%
, Morales%
\BCBL {}\ \BBA {} Santamar{\'\i}a%
}{%
Gonz{\'a}lez-Manteiga%
\ \protect \BOthers {.}}{%
{\protect \APACyear {2008}}%
}]{%
Gonzalez_etal2008}
\APACinsertmetastar {%
Gonzalez_etal2008}%
\begin{APACrefauthors}%
Gonz{\'a}lez-Manteiga, W.%
, Lombard{\'\i}a, M\BPBI J.%
, Molina, I.%
, Morales, D.%
\BCBL {}\ \BBA {} Santamar{\'\i}a, L.%
\end{APACrefauthors}%
\unskip\
\newblock
\APACrefYearMonthDay{2008}{}{}.
\newblock
{\BBOQ}\APACrefatitle {{Bootstrap Mean Squared Error of a Small-Area EBLUP}}
  {{Bootstrap Mean Squared Error of a Small-Area EBLUP}}.{\BBCQ}
\newblock
\APACjournalVolNumPages{Journal of Statistical Computation and
  Simulation}{78}{5}{443--462}.
\PrintBackRefs{\CurrentBib}

\bibitem [\protect \citeauthoryear {%
Graf%
, Mar{\'i}n%
\BCBL {}\ \BBA {} Molina%
}{%
Graf%
\ \protect \BOthers {.}}{%
{\protect \APACyear {2019}}%
}]{%
Graf_etal2019}
\APACinsertmetastar {%
Graf_etal2019}%
\begin{APACrefauthors}%
Graf, M.%
, Mar{\'i}n, J\BPBI M.%
\BCBL {}\ \BBA {} Molina, I.%
\end{APACrefauthors}%
\unskip\
\newblock
\APACrefYearMonthDay{2019}{}{}.
\newblock
{\BBOQ}\APACrefatitle {{A Generalized Mixed Model for Skewed Distributions
  Applied to Small Area Estimation}} {{A Generalized Mixed Model for Skewed
  Distributions Applied to Small Area Estimation}}.{\BBCQ}
\newblock
\APACjournalVolNumPages{Test}{28}{2}{565--597}.
\PrintBackRefs{\CurrentBib}

\bibitem [\protect \citeauthoryear {%
Greenwell%
}{%
Greenwell%
}{%
{\protect \APACyear {2017}}%
}]{%
Greenwell_2017}
\APACinsertmetastar {%
Greenwell_2017}%
\begin{APACrefauthors}%
Greenwell, B\BPBI M.%
\end{APACrefauthors}%
\unskip\
\newblock
\APACrefYearMonthDay{2017}{}{}.
\newblock
{\BBOQ}\APACrefatitle {{pdp: An R Package for Constructing Partial Dependence
  Plots}} {{pdp: An R Package for Constructing Partial Dependence
  Plots}}.{\BBCQ}
\newblock
\APACjournalVolNumPages{{The R Journal}}{9}{1}{421--436}.
\PrintBackRefs{\CurrentBib}

\bibitem [\protect \citeauthoryear {%
Greenwell%
, Boehmke%
\BCBL {}\ \BBA {} Gray%
}{%
Greenwell%
\ \protect \BOthers {.}}{%
{\protect \APACyear {2020}}%
}]{%
Greenwell_etal2020}
\APACinsertmetastar {%
Greenwell_etal2020}%
\begin{APACrefauthors}%
Greenwell, B\BPBI M.%
, Boehmke, B.%
\BCBL {}\ \BBA {} Gray, B.%
\end{APACrefauthors}%
\unskip\
\newblock
\APACrefYearMonthDay{2020}{}{}.
\newblock
{\BBOQ}\APACrefatitle {{Variable Importance Plots—An Introduction to the vip
  Package}} {{Variable Importance Plots—An Introduction to the vip
  Package}}.{\BBCQ}
\newblock
\APACjournalVolNumPages{The R Journal}{12}{1}{343--366}.
\PrintBackRefs{\CurrentBib}

\bibitem [\protect \citeauthoryear {%
Hajjem%
, Bellavance%
\BCBL {}\ \BBA {} Larocque%
}{%
Hajjem%
\ \protect \BOthers {.}}{%
{\protect \APACyear {2011}}%
}]{%
Hajjem_etal2011}
\APACinsertmetastar {%
Hajjem_etal2011}%
\begin{APACrefauthors}%
Hajjem, A.%
, Bellavance, F.%
\BCBL {}\ \BBA {} Larocque, D.%
\end{APACrefauthors}%
\unskip\
\newblock
\APACrefYearMonthDay{2011}{}{}.
\newblock
{\BBOQ}\APACrefatitle {{Mixed Effects Regression Trees for Clustered Data}}
  {{Mixed Effects Regression Trees for Clustered Data}}.{\BBCQ}
\newblock
\APACjournalVolNumPages{Statistics \& Probability Letters}{81}{4}{451--459}.
\PrintBackRefs{\CurrentBib}

\bibitem [\protect \citeauthoryear {%
Hajjem%
, Bellavance%
\BCBL {}\ \BBA {} Larocque%
}{%
Hajjem%
\ \protect \BOthers {.}}{%
{\protect \APACyear {2014}}%
}]{%
Hajjem2014}
\APACinsertmetastar {%
Hajjem2014}%
\begin{APACrefauthors}%
Hajjem, A.%
, Bellavance, F.%
\BCBL {}\ \BBA {} Larocque, D.%
\end{APACrefauthors}%
\unskip\
\newblock
\APACrefYearMonthDay{2014}{}{}.
\newblock
{\BBOQ}\APACrefatitle {{Mixed-Effects Random Forest for Clustered Data}}
  {{Mixed-Effects Random Forest for Clustered Data}}.{\BBCQ}
\newblock
\APACjournalVolNumPages{Journal of Statistical Computation and
  Simulation}{84}{6}{1313--1328}.
\PrintBackRefs{\CurrentBib}

\bibitem [\protect \citeauthoryear {%
Hall%
\ \BBA {} Maiti%
}{%
Hall%
\ \BBA {} Maiti%
}{%
{\protect \APACyear {2006}}%
}]{%
Hall_Maiti2006}
\APACinsertmetastar {%
Hall_Maiti2006}%
\begin{APACrefauthors}%
Hall, P.%
\BCBT {}\ \BBA {} Maiti, T.%
\end{APACrefauthors}%
\unskip\
\newblock
\APACrefYearMonthDay{2006}{}{}.
\newblock
{\BBOQ}\APACrefatitle {{On Parametric Bootstrap Methods for Small Area
  Prediction}} {{On Parametric Bootstrap Methods for Small Area
  Prediction}}.{\BBCQ}
\newblock
\APACjournalVolNumPages{Journal of the Royal Statistical Society: Series B
  (Statistical Methodology)}{68}{2}{221--238}.
\PrintBackRefs{\CurrentBib}

\bibitem [\protect \citeauthoryear {%
Hastie%
, Tibshirani%
\BCBL {}\ \BBA {} Friedman%
}{%
Hastie%
\ \protect \BOthers {.}}{%
{\protect \APACyear {2009}}%
}]{%
Hastie_etal2009}
\APACinsertmetastar {%
Hastie_etal2009}%
\begin{APACrefauthors}%
Hastie, T.%
, Tibshirani, R.%
\BCBL {}\ \BBA {} Friedman, J.%
\end{APACrefauthors}%
\unskip\
\newblock
\APACrefYear{2009}.
\newblock
\APACrefbtitle {{The Elements of Statistical Learning: Data Mining, Inference,
  and Prediction}} {{The Elements of Statistical Learning: Data Mining,
  Inference, and Prediction}}.
\newblock
\APACaddressPublisher{}{Springer Science \& Business Media}.
\PrintBackRefs{\CurrentBib}

\bibitem [\protect \citeauthoryear {%
Jiang%
\ \BBA {} Rao%
}{%
Jiang%
\ \BBA {} Rao%
}{%
{\protect \APACyear {2020}}%
}]{%
JiangRao2020}
\APACinsertmetastar {%
JiangRao2020}%
\begin{APACrefauthors}%
Jiang, J.%
\BCBT {}\ \BBA {} Rao, J\BPBI S.%
\end{APACrefauthors}%
\unskip\
\newblock
\APACrefYearMonthDay{2020}{}{}.
\newblock
{\BBOQ}\APACrefatitle {{Robust Small Area Estimation: An Overview}} {{Robust
  Small Area Estimation: An Overview}}.{\BBCQ}
\newblock
\APACjournalVolNumPages{Annual Review of Statistics and its
  Application}{7}{1}{337--360}.
\PrintBackRefs{\CurrentBib}

\bibitem [\protect \citeauthoryear {%
Kreutzmann%
\ \protect \BOthers {.}}{%
Kreutzmann%
\ \protect \BOthers {.}}{%
{\protect \APACyear {2019}}%
}]{%
Kreutzmann_etal2019}
\APACinsertmetastar {%
Kreutzmann_etal2019}%
\begin{APACrefauthors}%
Kreutzmann, A\BHBI K.%
, Pannier, S.%
, Rojas-Perilla, N.%
, Schmid, T.%
, Templ, M.%
\BCBL {}\ \BBA {} Tzavidis, N.%
\end{APACrefauthors}%
\unskip\
\newblock
\APACrefYearMonthDay{2019}{}{}.
\newblock
{\BBOQ}\APACrefatitle {{The R Package emdi for Estimating and Mapping
  Regionally Disaggregated Indicators}} {{The R Package emdi for Estimating and
  Mapping Regionally Disaggregated Indicators}}.{\BBCQ}
\newblock
\APACjournalVolNumPages{Journal of Statistical Software}{91}{7}{1--33}.
\PrintBackRefs{\CurrentBib}

\bibitem [\protect \citeauthoryear {%
Lambert%
\ \BBA {} Park%
}{%
Lambert%
\ \BBA {} Park%
}{%
{\protect \APACyear {2019}}%
}]{%
lambert_hyunmin2019}
\APACinsertmetastar {%
lambert_hyunmin2019}%
\begin{APACrefauthors}%
Lambert, F.%
\BCBT {}\ \BBA {} Park, H.%
\end{APACrefauthors}%
\unskip\
\newblock
\APACrefYearMonthDay{2019}{}{}.
\newblock
{\BBOQ}\APACrefatitle {{Income Inequality and Government Transfers in Mexico}}
  {{Income Inequality and Government Transfers in Mexico}}.{\BBCQ}
\newblock
\APACjournalVolNumPages{IMF Working Papers}{}{148}{}.
\PrintBackRefs{\CurrentBib}

\bibitem [\protect \citeauthoryear {%
Liaw%
\ \BBA {} Wiener%
}{%
Liaw%
\ \BBA {} Wiener%
}{%
{\protect \APACyear {2002}}%
}]{%
Liaw_Wiener2002}
\APACinsertmetastar {%
Liaw_Wiener2002}%
\begin{APACrefauthors}%
Liaw, A.%
\BCBT {}\ \BBA {} Wiener, M.%
\end{APACrefauthors}%
\unskip\
\newblock
\APACrefYearMonthDay{2002}{}{}.
\newblock
{\BBOQ}\APACrefatitle {{Classification and Regression by randomForest}}
  {{Classification and Regression by randomForest}}.{\BBCQ}
\newblock
\APACjournalVolNumPages{R News}{2}{3}{18-22}.
\PrintBackRefs{\CurrentBib}

\bibitem [\protect \citeauthoryear {%
Marchetti%
\ \protect \BOthers {.}}{%
Marchetti%
\ \protect \BOthers {.}}{%
{\protect \APACyear {2015}}%
}]{%
Marchetti_etal2015}
\APACinsertmetastar {%
Marchetti_etal2015}%
\begin{APACrefauthors}%
Marchetti, S.%
, Giusti, C.%
, Pratesi, M.%
, Salvati, N.%
, Giannotti, F.%
, Pedreschi, D.%
\BDBL {}Gabrielli, L.%
\end{APACrefauthors}%
\unskip\
\newblock
\APACrefYearMonthDay{2015}{}{}.
\newblock
{\BBOQ}\APACrefatitle {Small Area Model-Based Estimators Using Big Data
  Sources} {Small area model-based estimators using big data sources}.{\BBCQ}
\newblock
\APACjournalVolNumPages{Journal of Official Statistics}{31}{2}{263--281}.
\PrintBackRefs{\CurrentBib}

\bibitem [\protect \citeauthoryear {%
Marchetti%
\ \BBA {} Tzavidis%
}{%
Marchetti%
\ \BBA {} Tzavidis%
}{%
{\protect \APACyear {2021}}%
}]{%
Marchetti_Tzavidis2021}
\APACinsertmetastar {%
Marchetti_Tzavidis2021}%
\begin{APACrefauthors}%
Marchetti, S.%
\BCBT {}\ \BBA {} Tzavidis, N.%
\end{APACrefauthors}%
\unskip\
\newblock
\APACrefYearMonthDay{2021}{}{}.
\newblock
{\BBOQ}\APACrefatitle {Robust Estimation of the Theil Index and the Gini
  Coeffient for Small Areas} {Robust estimation of the theil index and the gini
  coeffient for small areas}.{\BBCQ}
\newblock
\APACjournalVolNumPages{Journal of Official Statistics}{37}{4}{955--979}.
\PrintBackRefs{\CurrentBib}

\bibitem [\protect \citeauthoryear {%
Marino%
, Ranalli%
, Salvati%
\BCBL {}\ \BBA {} Alfo%
}{%
Marino%
\ \protect \BOthers {.}}{%
{\protect \APACyear {2019}}%
}]{%
Marino2019}
\APACinsertmetastar {%
Marino2019}%
\begin{APACrefauthors}%
Marino, M\BPBI F.%
, Ranalli, M\BPBI G.%
, Salvati, N.%
\BCBL {}\ \BBA {} Alfo, M.%
\end{APACrefauthors}%
\unskip\
\newblock
\APACrefYearMonthDay{2019}{}{}.
\newblock
{\BBOQ}\APACrefatitle {{Semi-Parametric Empirical Best Prediction for Small
  Area Estimation of Unemployment Indicators}} {{Semi-Parametric Empirical Best
  Prediction for Small Area Estimation of Unemployment Indicators}}.{\BBCQ}
\newblock
\APACjournalVolNumPages{Annals of Applied Statistics}{}{}{forthcoming}.
\PrintBackRefs{\CurrentBib}

\bibitem [\protect \citeauthoryear {%
Marino%
, Tzavidis%
\BCBL {}\ \BBA {} Alfo%
}{%
Marino%
\ \protect \BOthers {.}}{%
{\protect \APACyear {2018}}%
}]{%
Marino2016}
\APACinsertmetastar {%
Marino2016}%
\begin{APACrefauthors}%
Marino, M\BPBI F.%
, Tzavidis, N.%
\BCBL {}\ \BBA {} Alfo, M.%
\end{APACrefauthors}%
\unskip\
\newblock
\APACrefYearMonthDay{2018}{}{}.
\newblock
{\BBOQ}\APACrefatitle {{Mixed hidden Markov Quantile Regression Models for
  Longitudinal Data with Possibly Incomplete Sequences}} {{Mixed hidden Markov
  Quantile Regression Models for Longitudinal Data with Possibly Incomplete
  Sequences}}.{\BBCQ}
\newblock
\APACjournalVolNumPages{Statistical Methods in Medical
  Research}{27}{7}{2231--2246}.
\PrintBackRefs{\CurrentBib}

\bibitem [\protect \citeauthoryear {%
McConville%
\ \BBA {} Toth%
}{%
McConville%
\ \BBA {} Toth%
}{%
{\protect \APACyear {2019}}%
}]{%
MoConvilleetal2019}
\APACinsertmetastar {%
MoConvilleetal2019}%
\begin{APACrefauthors}%
McConville, K\BPBI S.%
\BCBT {}\ \BBA {} Toth, D.%
\end{APACrefauthors}%
\unskip\
\newblock
\APACrefYearMonthDay{2019}{}{}.
\newblock
{\BBOQ}\APACrefatitle {{Automated Selection of Post-Strata using a
  Model-Assisted Regression Tree Estimator}} {{Automated Selection of
  Post-Strata using a Model-Assisted Regression Tree Estimator}}.{\BBCQ}
\newblock
\APACjournalVolNumPages{Scandinavian Journal of Statistics}{46}{2}{389--413}.
\PrintBackRefs{\CurrentBib}

\bibitem [\protect \citeauthoryear {%
Mendez%
}{%
Mendez%
}{%
{\protect \APACyear {2008}}%
}]{%
Mendez2008}
\APACinsertmetastar {%
Mendez2008}%
\begin{APACrefauthors}%
Mendez, G.%
\end{APACrefauthors}%
\unskip\
\newblock
\APACrefYear{2008}.
\unskip\
\newblock
\APACrefbtitle {{Tree-Based Mehtods to Model Dependent Data}} {{Tree-Based
  Mehtods to Model Dependent Data}}\ \APACtypeAddressSchool {\BUPhD}{}{}.
\unskip\
\newblock
\APACaddressSchool {}{Arizona State University}.
\PrintBackRefs{\CurrentBib}

\bibitem [\protect \citeauthoryear {%
Mendez%
\ \BBA {} Lohr%
}{%
Mendez%
\ \BBA {} Lohr%
}{%
{\protect \APACyear {2011}}%
}]{%
Mendez_Lohr2011}
\APACinsertmetastar {%
Mendez_Lohr2011}%
\begin{APACrefauthors}%
Mendez, G.%
\BCBT {}\ \BBA {} Lohr, S.%
\end{APACrefauthors}%
\unskip\
\newblock
\APACrefYearMonthDay{2011}{}{}.
\newblock
{\BBOQ}\APACrefatitle {{Estimating Residual Variance in Random Forest
  Regression}} {{Estimating Residual Variance in Random Forest
  Regression}}.{\BBCQ}
\newblock
\APACjournalVolNumPages{Computational Statistics \& Data
  Analysis}{55}{11}{2937--2950}.
\PrintBackRefs{\CurrentBib}

\bibitem [\protect \citeauthoryear {%
Molina%
\ \BBA {} Marhuenda%
}{%
Molina%
\ \BBA {} Marhuenda%
}{%
{\protect \APACyear {2015}}%
}]{%
Molina_Marhuenda:2015}
\APACinsertmetastar {%
Molina_Marhuenda:2015}%
\begin{APACrefauthors}%
Molina, I.%
\BCBT {}\ \BBA {} Marhuenda, Y.%
\end{APACrefauthors}%
\unskip\
\newblock
\APACrefYearMonthDay{2015}{}{}.
\newblock
{\BBOQ}\APACrefatitle {{{sae}: An {R} Package for Small Area Estimation}}
  {{{sae}: An {R} Package for Small Area Estimation}}.{\BBCQ}
\newblock
\APACjournalVolNumPages{The R Journal}{7}{1}{81--98}.
\PrintBackRefs{\CurrentBib}

\bibitem [\protect \citeauthoryear {%
Molina%
\ \BBA {} Rao%
}{%
Molina%
\ \BBA {} Rao%
}{%
{\protect \APACyear {2010}}%
}]{%
Molina_rao2010}
\APACinsertmetastar {%
Molina_rao2010}%
\begin{APACrefauthors}%
Molina, I.%
\BCBT {}\ \BBA {} Rao, J\BPBI N\BPBI K.%
\end{APACrefauthors}%
\unskip\
\newblock
\APACrefYearMonthDay{2010}{}{}.
\newblock
{\BBOQ}\APACrefatitle {{Small Area Estimation of Poverty Indicators}} {{Small
  Area Estimation of Poverty Indicators}}.{\BBCQ}
\newblock
\APACjournalVolNumPages{Canadian Journal of Statistics}{38}{3}{369--385}.
\PrintBackRefs{\CurrentBib}

\bibitem [\protect \citeauthoryear {%
OECD%
}{%
OECD%
}{%
{\protect \APACyear {2021}}%
}]{%
Oecd_21}
\APACinsertmetastar {%
Oecd_21}%
\begin{APACrefauthors}%
OECD.%
\end{APACrefauthors}%
\unskip\
\newblock
\APACrefYearMonthDay{2021}{}{}.
\newblock
{\BBOQ}\APACrefatitle {Income distribution} {Income distribution}.{\BBCQ}
\newblock
\APACjournalVolNumPages{OECD Social and Welfare Statistics (database)}{}{}{}.
\newblock
\begin{APACrefURL} \url{https://doi.org/10.1787/data-00654-en} \end{APACrefURL}
\newblock
\APACrefnote{accessed on 13 April 2021}
\PrintBackRefs{\CurrentBib}

\bibitem [\protect \citeauthoryear {%
Opsomer%
, Claeskens%
, Ranalli%
, Kauermann%
\BCBL {}\ \BBA {} Breidt%
}{%
Opsomer%
\ \protect \BOthers {.}}{%
{\protect \APACyear {2008}}%
}]{%
Opsomeretal2008}
\APACinsertmetastar {%
Opsomeretal2008}%
\begin{APACrefauthors}%
Opsomer, J\BPBI D.%
, Claeskens, G.%
, Ranalli, M\BPBI G.%
, Kauermann, G.%
\BCBL {}\ \BBA {} Breidt, F.%
\end{APACrefauthors}%
\unskip\
\newblock
\APACrefYearMonthDay{2008}{}{}.
\newblock
{\BBOQ}\APACrefatitle {{Non-Parametric Small Area Estimation Using Penalized
  Spline Regression}} {{Non-Parametric Small Area Estimation Using Penalized
  Spline Regression}}.{\BBCQ}
\newblock
\APACjournalVolNumPages{Journal of the Royal Statistical Society: Series B
  (Statistical Methodology)}{70}{1}{265--286}.
\PrintBackRefs{\CurrentBib}

\bibitem [\protect \citeauthoryear {%
Prasad%
\ \BBA {} Rao%
}{%
Prasad%
\ \BBA {} Rao%
}{%
{\protect \APACyear {1990}}%
}]{%
Prasad_Rao1990}
\APACinsertmetastar {%
Prasad_Rao1990}%
\begin{APACrefauthors}%
Prasad, N\BPBI G\BPBI N.%
\BCBT {}\ \BBA {} Rao, J\BPBI N\BPBI K.%
\end{APACrefauthors}%
\unskip\
\newblock
\APACrefYearMonthDay{1990}{}{}.
\newblock
{\BBOQ}\APACrefatitle {{The Estimation of the Mean Squared Error of Small-Area
  Estimators}} {{The Estimation of the Mean Squared Error of Small-Area
  Estimators}}.{\BBCQ}
\newblock
\APACjournalVolNumPages{Journal of the American Statistical
  Association}{85}{409}{163--171}.
\PrintBackRefs{\CurrentBib}

\bibitem [\protect \citeauthoryear {%
{R Core Team}%
}{%
{R Core Team}%
}{%
{\protect \APACyear {2021}}%
}]{%
R_language}
\APACinsertmetastar {%
R_language}%
\begin{APACrefauthors}%
{R Core Team}.%
\end{APACrefauthors}%
\unskip\
\newblock
\APACrefYearMonthDay{2021}{}{}.
\newblock
{\BBOQ}\APACrefatitle {{R: A Language and Environment for Statistical
  Computing}} {{R: A Language and Environment for Statistical
  Computing}}{\BBCQ}\ [\bibcomputersoftwaremanual].
\newblock
\APACaddressPublisher{Vienna, Austria}{}.
\PrintBackRefs{\CurrentBib}

\bibitem [\protect \citeauthoryear {%
Rao%
\ \BBA {} Molina%
}{%
Rao%
\ \BBA {} Molina%
}{%
{\protect \APACyear {2015}}%
}]{%
Rao_Molina2015}
\APACinsertmetastar {%
Rao_Molina2015}%
\begin{APACrefauthors}%
Rao, J\BPBI N\BPBI K.%
\BCBT {}\ \BBA {} Molina, I.%
\end{APACrefauthors}%
\unskip\
\newblock
\APACrefYear{2015}.
\newblock
\APACrefbtitle {{Small Area Estimation}} {{Small Area Estimation}}\
  (\PrintOrdinal{2.nd}\ \BEd).
\newblock
\APACaddressPublisher{New Jersey: Wiley}{Wiley series in survey methodology}.
\PrintBackRefs{\CurrentBib}

\bibitem [\protect \citeauthoryear {%
Rojas-Perilla%
, Pannier%
, Schmid%
\BCBL {}\ \BBA {} Tzavidis%
}{%
Rojas-Perilla%
\ \protect \BOthers {.}}{%
{\protect \APACyear {2020}}%
}]{%
Rojas_etal2019}
\APACinsertmetastar {%
Rojas_etal2019}%
\begin{APACrefauthors}%
Rojas-Perilla, N.%
, Pannier, S.%
, Schmid, T.%
\BCBL {}\ \BBA {} Tzavidis, N.%
\end{APACrefauthors}%
\unskip\
\newblock
\APACrefYearMonthDay{2020}{}{}.
\newblock
{\BBOQ}\APACrefatitle {{Data-Driven Transformations in Small Area Estimation}}
  {{Data-Driven Transformations in Small Area Estimation}}.{\BBCQ}
\newblock
\APACjournalVolNumPages{Journal of the Royal Statistical Society: Series A
  (Statistics in Society)}{183}{1}{121--148}.
\PrintBackRefs{\CurrentBib}

\bibitem [\protect \citeauthoryear {%
Schmid%
, Bruckschen%
, Salvati%
\BCBL {}\ \BBA {} Zbiranski%
}{%
Schmid%
\ \protect \BOthers {.}}{%
{\protect \APACyear {2017}}%
}]{%
Schmid2017}
\APACinsertmetastar {%
Schmid2017}%
\begin{APACrefauthors}%
Schmid, T.%
, Bruckschen, F.%
, Salvati, N.%
\BCBL {}\ \BBA {} Zbiranski, T.%
\end{APACrefauthors}%
\unskip\
\newblock
\APACrefYearMonthDay{2017}{}{}.
\newblock
{\BBOQ}\APACrefatitle {Constructing sociodemographic indicators for national
  statistical institutes by using mobile phone data: estimating literacy rates
  in Senegal} {Constructing sociodemographic indicators for national
  statistical institutes by using mobile phone data: estimating literacy rates
  in senegal}.{\BBCQ}
\newblock
\APACjournalVolNumPages{Journal of the Royal Statistical Society: Series
  A}{180}{4}{1163-1190}.
\PrintBackRefs{\CurrentBib}

\bibitem [\protect \citeauthoryear {%
Scornet%
, Biau%
\BCBL {}\ \BBA {} Vert%
}{%
Scornet%
\ \protect \BOthers {.}}{%
{\protect \APACyear {2015}}%
}]{%
Scornet_etal2015}
\APACinsertmetastar {%
Scornet_etal2015}%
\begin{APACrefauthors}%
Scornet, E.%
, Biau, G.%
\BCBL {}\ \BBA {} Vert, J\BHBI P.%
\end{APACrefauthors}%
\unskip\
\newblock
\APACrefYearMonthDay{2015}{}{}.
\newblock
{\BBOQ}\APACrefatitle {{Consistency of Random Forests}} {{Consistency of Random
  Forests}}.{\BBCQ}
\newblock
\APACjournalVolNumPages{The Annals of Statistics}{43}{4}{1716--1741}.
\PrintBackRefs{\CurrentBib}

\bibitem [\protect \citeauthoryear {%
Sela%
\ \BBA {} Simonoff%
}{%
Sela%
\ \BBA {} Simonoff%
}{%
{\protect \APACyear {2012}}%
}]{%
Sela_Simonoff2012}
\APACinsertmetastar {%
Sela_Simonoff2012}%
\begin{APACrefauthors}%
Sela, R\BPBI J.%
\BCBT {}\ \BBA {} Simonoff, J\BPBI S.%
\end{APACrefauthors}%
\unskip\
\newblock
\APACrefYearMonthDay{2012}{}{}.
\newblock
{\BBOQ}\APACrefatitle {{RE-EM Trees: A Data Mining Approach for Longitudinal
  and Clustered Data}} {{RE-EM Trees: A Data Mining Approach for Longitudinal
  and Clustered Data}}.{\BBCQ}
\newblock
\APACjournalVolNumPages{Machine Learning}{86}{2}{169--207}.
\PrintBackRefs{\CurrentBib}

\bibitem [\protect \citeauthoryear {%
Sexton%
\ \BBA {} Laake%
}{%
Sexton%
\ \BBA {} Laake%
}{%
{\protect \APACyear {2009}}%
}]{%
Sexton_Laake2009}
\APACinsertmetastar {%
Sexton_Laake2009}%
\begin{APACrefauthors}%
Sexton, J.%
\BCBT {}\ \BBA {} Laake, P.%
\end{APACrefauthors}%
\unskip\
\newblock
\APACrefYearMonthDay{2009}{}{}.
\newblock
{\BBOQ}\APACrefatitle {{Standard Errors for Bagged and Random Forest
  Estimators}} {{Standard Errors for Bagged and Random Forest
  Estimators}}.{\BBCQ}
\newblock
\APACjournalVolNumPages{Computational Statistics \& Data
  Analysis}{53}{3}{801--811}.
\PrintBackRefs{\CurrentBib}

\bibitem [\protect \citeauthoryear {%
Smits%
\ \BBA {} Permanyer%
}{%
Smits%
\ \BBA {} Permanyer%
}{%
{\protect \APACyear {2019}}%
}]{%
smits_Permanyer2019}
\APACinsertmetastar {%
smits_Permanyer2019}%
\begin{APACrefauthors}%
Smits, J.%
\BCBT {}\ \BBA {} Permanyer, I.%
\end{APACrefauthors}%
\unskip\
\newblock
\APACrefYearMonthDay{2019}{}{}.
\newblock
{\BBOQ}\APACrefatitle {{The Subnational Human Development Database}} {{The
  Subnational Human Development Database}}.{\BBCQ}
\newblock
\APACjournalVolNumPages{Scientific Database}{6}{190038}{}.
\PrintBackRefs{\CurrentBib}

\bibitem [\protect \citeauthoryear {%
Sugasawa%
\ \BBA {} Kubokawa%
}{%
Sugasawa%
\ \BBA {} Kubokawa%
}{%
{\protect \APACyear {2017}}%
}]{%
sugasawa2017transforming}
\APACinsertmetastar {%
sugasawa2017transforming}%
\begin{APACrefauthors}%
Sugasawa, S.%
\BCBT {}\ \BBA {} Kubokawa, T.%
\end{APACrefauthors}%
\unskip\
\newblock
\APACrefYearMonthDay{2017}{}{}.
\newblock
{\BBOQ}\APACrefatitle {{Transforming Response Values in Small Area Prediction}}
  {{Transforming Response Values in Small Area Prediction}}.{\BBCQ}
\newblock
\APACjournalVolNumPages{Computational Statistics \& Data
  Analysis}{114}{}{47--60}.
\PrintBackRefs{\CurrentBib}

\bibitem [\protect \citeauthoryear {%
Sugasawa%
\ \BBA {} Kubokawa%
}{%
Sugasawa%
\ \BBA {} Kubokawa%
}{%
{\protect \APACyear {2019}}%
}]{%
sugasawa2019adaptively}
\APACinsertmetastar {%
sugasawa2019adaptively}%
\begin{APACrefauthors}%
Sugasawa, S.%
\BCBT {}\ \BBA {} Kubokawa, T.%
\end{APACrefauthors}%
\unskip\
\newblock
\APACrefYearMonthDay{2019}{}{}.
\newblock
{\BBOQ}\APACrefatitle {{Adaptively Transformed Mixed-Model Prediction of
  General Finite-Population Parameters}} {{Adaptively Transformed Mixed-Model
  Prediction of General Finite-Population Parameters}}.{\BBCQ}
\newblock
\APACjournalVolNumPages{Scandinavian Journal of Statistics}{46}{4}{1025--1046}.
\PrintBackRefs{\CurrentBib}

\bibitem [\protect \citeauthoryear {%
Tzavidis%
, Marchetti%
\BCBL {}\ \BBA {} Chambers%
}{%
Tzavidis%
\ \protect \BOthers {.}}{%
{\protect \APACyear {2010}}%
}]{%
Tzavidis_etal2010}
\APACinsertmetastar {%
Tzavidis_etal2010}%
\begin{APACrefauthors}%
Tzavidis, N.%
, Marchetti, S.%
\BCBL {}\ \BBA {} Chambers, R.%
\end{APACrefauthors}%
\unskip\
\newblock
\APACrefYearMonthDay{2010}{}{}.
\newblock
{\BBOQ}\APACrefatitle {ROBUST ESTIMATION OF SMALL-AREA MEANS AND QUANTILES}
  {Robust estimation of small-area means and quantiles}.{\BBCQ}
\newblock
\APACjournalVolNumPages{Australian \& New Zealand Journal of
  Statistics}{52}{2}{167-186}.
\PrintBackRefs{\CurrentBib}

\bibitem [\protect \citeauthoryear {%
Tzavidis%
, Zhang%
, Luna%
, Schmid%
\BCBL {}\ \BBA {} Rojas-Perilla%
}{%
Tzavidis%
\ \protect \BOthers {.}}{%
{\protect \APACyear {2018}}%
}]{%
Tzavidis_etal2018}
\APACinsertmetastar {%
Tzavidis_etal2018}%
\begin{APACrefauthors}%
Tzavidis, N.%
, Zhang, L\BHBI C.%
, Luna, A.%
, Schmid, T.%
\BCBL {}\ \BBA {} Rojas-Perilla, N.%
\end{APACrefauthors}%
\unskip\
\newblock
\APACrefYearMonthDay{2018}{}{}.
\newblock
{\BBOQ}\APACrefatitle {{From Start to Finish: A Framework for the Production of
  Small Area Official Statistics}} {{From Start to Finish: A Framework for the
  Production of Small Area Official Statistics}}.{\BBCQ}
\newblock
\APACjournalVolNumPages{Journal of the Royal Statistical Society: Series A
  (Statistics in Society)}{181}{4}{927--979}.
\PrintBackRefs{\CurrentBib}

\bibitem [\protect \citeauthoryear {%
Varian%
}{%
Varian%
}{%
{\protect \APACyear {2014}}%
}]{%
Varian2014}
\APACinsertmetastar {%
Varian2014}%
\begin{APACrefauthors}%
Varian, H\BPBI R.%
\end{APACrefauthors}%
\unskip\
\newblock
\APACrefYearMonthDay{2014}{}{}.
\newblock
{\BBOQ}\APACrefatitle {{Big Data: New Tricks for Econometrics}} {{Big Data: New
  Tricks for Econometrics}}.{\BBCQ}
\newblock
\APACjournalVolNumPages{Journal of Economic Perspectives}{28}{2}{3--28}.
\PrintBackRefs{\CurrentBib}

\bibitem [\protect \citeauthoryear {%
Wager%
\ \BBA {} Athey%
}{%
Wager%
\ \BBA {} Athey%
}{%
{\protect \APACyear {2018}}%
}]{%
Wager_Athey2018}
\APACinsertmetastar {%
Wager_Athey2018}%
\begin{APACrefauthors}%
Wager, S.%
\BCBT {}\ \BBA {} Athey, S.%
\end{APACrefauthors}%
\unskip\
\newblock
\APACrefYearMonthDay{2018}{}{}.
\newblock
{\BBOQ}\APACrefatitle {{Estimation and Inference of Heterogeneous Treatment
  Effects using Random Forests}} {{Estimation and Inference of Heterogeneous
  Treatment Effects using Random Forests}}.{\BBCQ}
\newblock
\APACjournalVolNumPages{Journal of the American Statistical
  Association}{113}{523}{1228--1242}.
\PrintBackRefs{\CurrentBib}

\bibitem [\protect \citeauthoryear {%
Wager%
, Hastie%
\BCBL {}\ \BBA {} Efron%
}{%
Wager%
\ \protect \BOthers {.}}{%
{\protect \APACyear {2014}}%
}]{%
wager_etal2014}
\APACinsertmetastar {%
wager_etal2014}%
\begin{APACrefauthors}%
Wager, S.%
, Hastie, T.%
\BCBL {}\ \BBA {} Efron, B.%
\end{APACrefauthors}%
\unskip\
\newblock
\APACrefYearMonthDay{2014}{}{}.
\newblock
{\BBOQ}\APACrefatitle {{Confidence Intervals for Random Forests: The Jackknife
  and the Infinitesimal Jackknife}} {{Confidence Intervals for Random Forests:
  The Jackknife and the Infinitesimal Jackknife}}.{\BBCQ}
\newblock
\APACjournalVolNumPages{The Journal of Machine Learning
  Research}{15}{1}{1625--1651}.
\PrintBackRefs{\CurrentBib}

\bibitem [\protect \citeauthoryear {%
Winham%
, Freimuth%
\BCBL {}\ \BBA {} Biernacka%
}{%
Winham%
\ \protect \BOthers {.}}{%
{\protect \APACyear {2013}}%
}]{%
Winham_etal2013}
\APACinsertmetastar {%
Winham_etal2013}%
\begin{APACrefauthors}%
Winham, S\BPBI J.%
, Freimuth, R\BPBI R.%
\BCBL {}\ \BBA {} Biernacka, J\BPBI M.%
\end{APACrefauthors}%
\unskip\
\newblock
\APACrefYearMonthDay{2013}{}{}.
\newblock
{\BBOQ}\APACrefatitle {{A Weighted Random Forests Approach to Improve
  Predictive Performance}} {{A Weighted Random Forests Approach to Improve
  Predictive Performance}}.{\BBCQ}
\newblock
\APACjournalVolNumPages{Statistical Analysis and Data Mining}{6}{6}{496--505}.
\PrintBackRefs{\CurrentBib}

\bibitem [\protect \citeauthoryear {%
Wood%
}{%
Wood%
}{%
{\protect \APACyear {2017}}%
}]{%
Wood_2017}
\APACinsertmetastar {%
Wood_2017}%
\begin{APACrefauthors}%
Wood, S.%
\end{APACrefauthors}%
\unskip\
\newblock
\APACrefYear{2017}.
\newblock
\APACrefbtitle {Generalized Additive Models: An Introduction with R}
  {Generalized additive models: An introduction with r}\ (\PrintOrdinal{2}\
  \BEd).
\newblock
\APACaddressPublisher{}{Chapman and Hall/CRC}.
\PrintBackRefs{\CurrentBib}

\bibitem [\protect \citeauthoryear {%
Wu%
\ \BBA {} Zhang%
}{%
Wu%
\ \BBA {} Zhang%
}{%
{\protect \APACyear {2006}}%
}]{%
Wu_Zhang2006}
\APACinsertmetastar {%
Wu_Zhang2006}%
\begin{APACrefauthors}%
Wu, H.%
\BCBT {}\ \BBA {} Zhang, J\BHBI T.%
\end{APACrefauthors}%
\unskip\
\newblock
\APACrefYear{2006}.
\newblock
\APACrefbtitle {{Nonparametric Regression Methods for Longitudinal Data
  Analysis: Mixed-Effects Modeling Approaches}} {{Nonparametric Regression
  Methods for Longitudinal Data Analysis: Mixed-Effects Modeling Approaches}}.
\newblock
\APACaddressPublisher{}{John Wiley \& Sons}.
\PrintBackRefs{\CurrentBib}

\bibitem [\protect \citeauthoryear {%
Zhang%
, Zimmerman%
, Nettleton%
\BCBL {}\ \BBA {} Nordman%
}{%
Zhang%
\ \protect \BOthers {.}}{%
{\protect \APACyear {2019}}%
}]{%
Zhang2019}
\APACinsertmetastar {%
Zhang2019}%
\begin{APACrefauthors}%
Zhang, H.%
, Zimmerman, J.%
, Nettleton, D.%
\BCBL {}\ \BBA {} Nordman, D\BPBI J.%
\end{APACrefauthors}%
\unskip\
\newblock
\APACrefYearMonthDay{2019}{}{}.
\newblock
{\BBOQ}\APACrefatitle {{Random Forest Prediction Intervals}} {{Random Forest
  Prediction Intervals}}.{\BBCQ}
\newblock
\APACjournalVolNumPages{The American Statistician}{74}{4}{392--406}.
\PrintBackRefs{\CurrentBib}

\end{thebibliography}

\end{document}